\documentclass[camera,letterpaper,nomarginnotes,nonarrowgutter]{jpaper}

\usepackage{amsmath,amssymb,amsfonts}
\usepackage{algorithmic}
\usepackage{graphicx}
\usepackage{textcomp}

\usepackage{tikz}
\usepackage{graphicx}
\usepackage{xcolor}
\usepackage{url}
\usepackage{cite}
\usepackage{stfloats}

\usepackage{fancyhdr}
\usepackage{listings}
\usepackage{comment}
\usepackage{subfigure}
\usepackage{subcaption}
\usepackage{pifont}
\usepackage{booktabs}
\usepackage{tabularx}
\usepackage{lipsum}
\usepackage[colorinlistoftodos,prependcaption,textsize=small]{todonotes} 

\usepackage{duckuments} 
\usepackage{filecontents}
\usepackage{multirow}
\usepackage{booktabs}
\usepackage{balance}
\usepackage{setspace}
\usepackage[linesnumbered]{algorithm2e}
\usepackage{siunitx} 
\usepackage{multicol}
\usepackage{threeparttable}
\usepackage{titlesec}
\usepackage{marginnote}
\usepackage{enumitem}
\usepackage[bookmarks=true,breaklinks=true,letterpaper=true,colorlinks,citecolor=black,linkcolor=blue,urlcolor=blue]{hyperref}
\usepackage{datetime}

\def\BibTeX{{\rm B\kern-.05em{\sc i\kern-.025em b}\kern-.08em
    T\kern-.1667em\lower.7ex\hbox{E}\kern-.125emX}}

\newif\ifrev
\newif\ifiterations
\iterationsfalse
\revfalse

\newcommand{\X}[0]{EasyDRAM}

\ifiterations
\newcommand{\atb}[2]{\ifnum#1=\value{version}\textcolor{orange}{#2}\else{#2}\fi}
\newcommand{\om}[2]{\ifnum#1=\value{version}\textcolor{blue}{#2}\else{#2}\fi}
\newcommand{\ous}[2]{\ifnum#1=\value{version}\textcolor{purple}{#2}\else{#2}\fi}
\newcommand{\dn}[2]{\ifnum#1=\value{version}\textcolor{magenta}{#2}\else{#2}\fi}

\newcommand{\atbcomment}[1]{\todo[size=\scriptsize, linecolor=orange, bordercolor=orange, backgroundcolor=white]{\textcolor{orange}{Atb:~#1}}}
\newcommand{\atbcrcomment}[2]{\ifnum#1=\value{version}\todo[size=\scriptsize, linecolor=orange, bordercolor=orange, backgroundcolor=white]{\textcolor{orange}{ATB:~#2}}\else{}\fi}
\newcommand{\omcomment}[2]{\ifnum#1=\value{version}\todo[size=\scriptsize, linecolor=blue, bordercolor=blue, backgroundcolor=white]{\textcolor{blue}{OM:~#2}}\else{}\fi}
\newcommand{\ouscomment}[1]{\todo[size=\scriptsize, linecolor=orange, bordercolor=orange, backgroundcolor=white]{\textcolor{purple}{Oguzhan:~#1}}}

\newcommand{\outline}[1]{\textcolor{orange}{OUTLINE:#1}}

\let\marginpar\marginnote
\paperwidth=\dimexpr \paperwidth + 5cm\relax
\oddsidemargin=\dimexpr\oddsidemargin + 2cm\relax
\evensidemargin=\dimexpr\evensidemargin + 2cm\relax
\marginparwidth=\dimexpr \marginparwidth + 2cm\relax

\else
\newcommand{\atb}[2]{#2}
\newcommand{\om}[2]{#2}
\newcommand{\ous}[2]{#2}
\newcommand{\dn}[2]{#2}

\newcommand{\atbcomment}[1]{}
\newcommand{\atbcrcomment}[2]{}
\newcommand{\omcomment}[2]{}
\newcommand{\ouscomment}[1]{}

\newcommand{\outline}[1]{}

\fi

\newcommand*\circled[1]{\tikz[baseline=(char.base)]{
            \node[shape=circle,fill,inner sep=2pt] (char) {\textcolor{white}{\scriptsize#1}};}}

\newcommand*\circledblue[1]{\tikz[baseline=(char.base)]{
            \node[shape=circle,fill=blue,inner sep=2pt] (char) {\textcolor{white}{\scriptsize#1}};}}

\newcommand*\circledteal[1]{\tikz[baseline=(char.base)]{
            \node[shape=circle,fill=teal,inner sep=2pt] (char) {\textcolor{white}{\scriptsize#1}};}}

\definecolor{timescalemotiv}{rgb}{0.62,0.19,0.52}

\newcommand*\circledres[1]{\tikz[baseline=(char.base)]{
            \node[shape=circle,fill=purple,inner sep=2pt] (char) {\textcolor{white}{\scriptsize#1}};}}
            
\newcommand*\circledmt[1]{\tikz[baseline=(char.base)]{
            \node[shape=circle,draw,inner sep=2pt] (char) {\textcolor{black}{\scriptsize#1}};}}

\DeclareRobustCommand\capcircledres[1]{\tikz[baseline=(char.base)]{
            \node[shape=circle,fill=purple,inner sep=2pt] (char) {\textcolor{white}{\scriptsize#1}};}}
            
\definecolor{dkgreen}{rgb}{0,0.45,0}
\definecolor{lightgray}{rgb}{0.92, 0.92, 0.92}
\definecolor{mauve}{rgb}{0.42,0,0.78}
\newcommand{\cmark}{\ding{51}}
\newcommand{\xmark}{\ding{55}}

\newcommand{\secref}[1]{§\ref{#1}}

\newcommand{\figref}[1]{Figure~\ref{#1}}

\newcommand{\head}[1]{\noindent\textbf{#1.}} 

\newcommand{\smc}{software memory controller}

\makeatletter
\g@addto@macro{\normalsize}{%
    \setlength{\abovedisplayskip}{2pt plus 0.5pt minus 1pt}
    \setlength{\belowdisplayskip}{2pt plus 0.5pt minus 1pt}
    \setlength{\abovedisplayshortskip}{0pt}
    \setlength{\belowdisplayshortskip}{0pt}
    \setlength{\intextsep}{2pt plus 1pt minus 1pt}
    \setlength{\textfloatsep}{2pt plus 1pt minus 1pt}
    \setlength{\skip\footins}{2pt plus 1pt minus 1pt}}
\makeatother

\definecolor{gfored}{rgb}{0.580, 0.050, 0.211}
\definecolor{ao}{rgb}{0.007, 0.520, 0.867}
\definecolor{moegi}{rgb}{0.357, 0.537, 0.188}
\definecolor{jl}{rgb}{1.0, 0.2, 0.8}
\definecolor{brown(web)}{rgb}{0.65, 0.16, 0.16}
\definecolor{bisque}{rgb}{1.0, 0.89, 0.77}
\definecolor{nbs}{rgb}{0.88, 0.07, 0.37}
\definecolor{yt}{rgb}{0.58, 0.44, 0.86}
\definecolor{iy}{rgb}{0.0, 0.36, 0.05}
\definecolor{burntorange}{rgb}{0.8, 0.33, 0.0}
\definecolor{lightmauve}{rgb}{0.86, 0.82, 1.0}
\definecolor{frenchblue}{rgb}{0.19, 0.55, 0.91}
\definecolor{amber}{rgb}{1.0, 0.49, 0.0}
\definecolor{awesome}{rgb}{1.0, 0.13, 0.32}
\definecolor{dollarbill}{rgb}{0.52,0.73,0.4}
\definecolor{moegi}{rgb}{0.357, 0.537, 0.188}
\definecolor{burgundy}{rgb}{0.5, 0.0, 0.13}
\definecolor{ballblue}{rgb}{0.13, 0.67, 0.8}
\definecolor{ups-truck}{rgb}{0.53, 0.28, 0.21}
\definecolor{airforceblue}{rgb}{0.36, 0.54, 0.66}
\definecolor{cadmiumgreen}{rgb}{0.0, 0.42, 0.24}
\definecolor{darkcyan}{rgb}{0.0, 0.55, 0.55}
\definecolor{caribbeangreen}{rgb}{0.0, 0.8, 0.6}
\definecolor{flamingopink}{rgb}{0.99, 0.56, 0.67}
\definecolor{jazzberryjam}{rgb}{0.65, 0.04, 0.37}
\definecolor{mediumpersianblue}{rgb}{0.0, 0.4, 0.65}
\definecolor{coolblack}{rgb}{0.0, 0.18, 0.39}
\definecolor{bleudefrance}{rgb}{0.19, 0.55, 0.91}
\definecolor{ao}{rgb}{0.0, 0.0, 1.0}
\definecolor{babyblueeyes}{rgb}{0.63, 0.79, 0.95}
\definecolor{darkwarmgray}{rgb}{0.2,0,0}
\definecolor{brightpink}{rgb}{1.0, 0.0, 0.5}
\definecolor{darkblue}{rgb}{0.0, 0.0, 0.55}

\ifrev
    \let\marginpar\marginnote
    \newcommand{\mirev}[1]{\textcolor{darkblue}{#1}}
    \newcommand{\cql}[2]{#2\todo[size=\small,color=jazzberryjam]{\textbf{\textrm{\textcolor{white}{#1}}}}}
    \newcommand{\iql}[2]{#2\todo[size=\small,color=babyblueeyes]{\textbf{\textrm{\textcolor{black}{#1}}}}}
    
\else

    \newcommand{\mirev}[1]{{#1}}
    \newcommand{\cql}[1]{}
    \newcommand{\iql}[1]{}
\fi

\newcounter{version}
\newcommand{\param}[1]{#1}

\newcommand{\versionnum}[0]{1.0}
\fancypagestyle{iterationsfirstpage}
{
    \fancyhead{}
    \fancyhead[C]{
    \textcolor{red}{CONFIDENTIAL DRAFT -- DO NOT DISTRIBUTE -- TO APPEAR IN DSN'25} \\ 
    \emph{\textcolor{blue}{Version \versionnum~---~\today, \ampmtime}}
    }
}

\makeatletter
\g@addto@macro{\normalsize}{%
  \setlength{\abovedisplayskip}{4pt plus 0.5pt minus 1pt}
  \setlength{\belowdisplayskip}{3pt plus 0.5pt minus 1pt}
  \setlength{\abovedisplayshortskip}{0pt}
  \setlength{\belowdisplayshortskip}{0pt}
  \setlength{\intextsep}{5pt plus 1pt minus 1pt}
  \setlength{\textfloatsep}{2pt plus 1pt minus 1pt}
  \setlength{\skip\footins}{5pt plus 1pt minus 1pt}
  \setlength{\abovecaptionskip}{2pt plus 0pt minus 0pt}}
\makeatother

\usepackage{titlesec}
\titlespacing\section{0pt}{5pt plus 1pt minus 1pt}{2pt plus 1pt minus 1pt}
\titlespacing\subsection{0pt}{5pt plus 1pt minus 1pt}{2pt plus 1pt minus 1pt}
\titlespacing\subsubsection{0pt}{5pt plus 1pt minus 1pt}{2pt plus 1pt minus 1pt}

\begin{document}
\bstctlcite{IEEEexample:BSTcontrol}

\title{\LARGE\X{}: \atb{2}{An FPGA-based Infrastructure for} Fast and Accurate\\\atb{2}{End-to-End} Evaluation of Emerging DRAM Techniques}

\newcommand{\affilETH}[0]{\textsuperscript{\S}}
\newcommand{\affilETU}[0]{\textsuperscript{$\dagger$}}
\newcommand{\affilUOS}[0]{\textsuperscript{$\ddagger$}}
\newcommand{\affilUOM}[0]{\textsuperscript{$\parallel$}}
\author{
{O\u{g}uzhan Canpolat\affilETU\affilETH}\qquad
{Ataberk Olgun\affilETH}\qquad
{David Novo\affilUOM}\qquad
{O\u{g}uz Ergin\affilUOS\affilETH\affilETU}\qquad
{Onur Mutlu\affilETH}\\
{\affilETH\emph{ETH Z{\"u}rich}}\qquad{}\affilETU\emph{TOBB ETÜ}\qquad{}\affilUOM\emph{LIRMM, Univ. Montpellier, CNRS}\qquad{}\affilUOS\emph{University of Sharjah}\vspace{-0.5em}
}

\maketitle

\ifiterations
\thispagestyle{iterationsfirstpage}
\else
\thispagestyle{plain}
\fi
\pagestyle{plain}
\pagenumbering{arabic}

\begin{abstract}

\mirev{DRAM \om{3}{is} a critical component of modern computing systems. Recent
works propose numerous techniques (that we call \emph{DRAM techniques}) to
enhance DRAM-based computing systems' throughput, reliability, and computing
capabilities (e.g., in-DRAM bulk data copy). Evaluating the system-wide benefits
of DRAM techniques is challenging as they often require modifications across
multiple layers of the \om{3}{computing} stack. Prior works propose FPGA-based
platforms for rapid end-to-end evaluation of DRAM techniques on real DRAM chips.
Unfortunately, existing platforms fall short in two \om{3}{major} aspects: (1)
they require deep expertise in hardware \atb{3}{description}
language\om{3}{s}, limiting accessibility; and (2) they are \emph{not} designed
to accurately model modern computing systems.}

We introduce \X{}, an FPGA-based framework for \mirev{rapid and accurate
end-to-end} evaluation of DRAM techniques on real DRAM chips. \mirev{\X{}
overcomes the main drawbacks of prior FPGA-based platforms with two key ideas.
First, \X{} removes the need for hardware description language expertise by
enabling developers to implement DRAM techniques using \om{3}{a} high-level
\om{3}{language} (C++). At runtime, \X{} executes the \om{3}{high-level}
software-defined memory system design in a programmable memory controller.
Second, \X{} tackles a fundamental challenge in accurately modeling modern
systems: real processors typically operate at significantly higher clock
frequencies than DRAM, a disparity that is difficult to replicate on FPGA
platforms. \X{} addresses this challenge by decoupling the processor–DRAM
interface and advancing the system state using a novel technique we call
\emph{time scaling}, which faithfully captures the timing behavior of the
modeled system.}

\mirev{We validate \X{}'s evaluation accuracy by comparing the memory latency
profile of a \om{3}{real CPU-based} system and its \om{3}{modeled}
implementation \om{4}{using} \X{}.} We demonstrate the ease of use of \X{} by evaluating
\param{two} DRAM techniques end-to-end \atb{0}{in a real \mirev{FPGA-based}}
system: \mirev{(1) in-DRAM bulk data copy (i.e., RowClone) and (2)
\om{3}{reduced-latency} DRAM access that exploits the latency variation across
DRAM cells.} \mirev{Implementing these two techniques requires no hardware
modifications and only \param{325} lines of C++ code over \X{}'s extensible code
base.}
We compare our results to prior \mirev{FPGA-based platforms}. \mirev{\X{} yields
more accurate results (e.g., by $\approx{}\!$20$\times{}$ for execution time)
than the state-of-the-art \om{3}{related} platform.} \mirev{We believe and hope that
\X{} \om{3}{will} enable innovative ideas in memory system design to rapidly come to
fruition.} \mirev{To aid future research, we open-source our \X{} implementation
at \url{https://github.com/CMU-SAFARI/EasyDRAM}.}

\end{abstract}
\section{Introduction}
\label{sec:introduction}

DRAM-based main memory is a critical component in modern computing systems. Many
prior works show that DRAM is increasingly becoming the primary bottleneck for
system performance and energy
efficiency~\cite{mutlu2013memory,mutlu2021modern,ghose2019processing,mutlu2025modernprimerprocessingmemory}
as the demand for higher performance, reduced latency, and increased capacity is
growing. To meet these requirements, many prior
works~\cite{kim2018solar,wang2018reducing,yaglikci2022hira,
luo2020clrdram,das2018vrldram,hassan2016chargecache,
lee2013tiered,lee2015adaptive,chang2016understanding,
chang2017thesis,choi2015multiple,chang2016lisa,
son2013reducing,yuksel2024simultaneous,yuksel2024functionally,olgun2021quactrng,
hassan2019crow,mathew2017using,koppula2019eden,orosa2023approximate,
chandrasekar2014exploiting,qin2023cdardram,zhang2016restore,shin2014nuat,
shin2015dram, talukder2019prelatpuf, kim2019drange, gao2022frac,
orosa2021codic,bostanci2022drstrange,gao2019computedram,kim2018dram}
introduce new \atb{2}{operations} \ous{1}{(we call these operations \emph{DRAM
techniques})} \mirev{that} violate timings set by the DRAM manufacturers to
improve the latency and computing capabilities of DRAM.

Evaluating the system-wide benefits of DRAM techniques is non-trivial as they
require modifications \mirev{across} multiple layers of the \mirev{comput\om{3}{ing}}
stack (e.g., memory controllers, memory mapping and allocation in the operating
system, and user applications), and require DRAM to be operated outside the
\mirev{current DRAM standards (e.g.,~\cite{jedec2017ddr4, jedec2020ddr5})}. For
example, it is possible to \atb{2}{reliably} copy \atb{2}{bulk data at} DRAM row
\atb{2}{granularity~\cite{seshadri2013rowclone} (typically 8 KiB)} by issuing row activation commands in
quick succession \atb{2}{in real DRAM
chips}~\cite{gao2019computedram,gao2022frac,olgun2021quactrng,olgun2022pidram,olgun2023drambender,yuksel2024functionally,yuksel2024simultaneous}.
\atb{2}{However, the DRAM} command sequence \atb{2}{to perform such copy
operations} is undefined in the current DRAM standards
(e.g.,~\cite{jedec2020ddr5,jedec2017ddr4,jedec2015lpddr4,jedec2020lpddr5}).
\atb{2}{Hence}, commercial \atb{2}{off-the-shelf} computing systems do
\emph{not} \atb{2}{perform those copy operations}.
\atb{3}{To alleviate the limitations of commercial systems in evaluating DRAM techniques,}
software simulators
\om{3}{(e.g.,~\cite{binkert2011gem5,kim2016ramulator,forlin2022sim2pim,power2014gem5,luo2023ramulator2})}
can be used to \atb{3}{rapidly model} 
\atb{3}{system support for} DRAM techniques. However, simulators have
\param{two} key drawbacks. First, correctly implementing the behavior of DRAM
techniques would require rigorous characterization studies and modeling efforts
by domain experts~\cite{kim2014flipping, kim2019drange,kim2018solar,
wang2018reducing,das2018vrldram,orosa2021deeper,
chang2017thesis,yaglikci2022hira,yaglikci2024spatial,
lee2015adaptive,yaglikci2022understanding,kim2018dram} \atb{2}{because a software
simulator does \emph{not} use a real DRAM chip}. Second, existing simulators
\mirev{exhibit} slow simulation speeds (e.g., \om{4}{only} a few
kHz~\cite{kim2016strober}), \ous{1}{thereby \mirev{requir\om{3}{ing}} a
\om{3}{large} amount of time (e.g., days or weeks) to \mirev{execute}
full-system workloads \mirev{end-to-end}}.

\mirev{To overcome the limitations of commercial systems and software simulators
in evaluating DRAM techniques, prior works~\cite{amid2020chipyard,
olgun2022pidram, hassan2017softmc, olgun2023drambender, biancolin2019fased,
karandikar2018firesim,mosanu2022pimulator,mosanu2023freezetime} propose
a variety of FPGA-based evaluation platforms that are broadly categorized into
three classes: (1) DRAM testing platforms, (2) FPGA-based simulators, and (3)
FPGA-based emulators. These platforms suffer from key drawbacks that prevent
their users from rapidly and accurately evaluating DRAM techniques end-to-end.} 

\noindent
\mirev{\textbf{Drawbacks of Existing FPGA-Based Platforms.}} First, DRAM testing
platforms \om{3}{(e.g.,~\cite{hassan2017softmc,olgun2023drambender})} allow DRAM
characterization studies to discover and test DRAM
techniques~\cite{kim2018solar,wang2018reducing,yaglikci2022hira,patel2017reaper,
luo2020clrdram,das2018vrldram,hassan2016chargecache,
lee2013tiered,lee2015adaptive,chang2016understanding,
chang2017thesis,choi2015multiple,chang2016lisa,
son2013reducing,yuksel2024simultaneous,yuksel2024functionally,olgun2021quactrng,
hassan2019crow,mathew2017using,koppula2019eden,orosa2023approximate,
chandrasekar2014exploiting,qin2023cdardram,zhang2016restore,shin2014nuat,
shin2015dram, talukder2019prelatpuf, kim2019drange, gao2022frac,
orosa2021codic,bostanci2022drstrange,gao2019computedram,kim2018dram}.
\atb{3}{These platforms are \emph{not} designed for system performance
evaluation. Thereby, they} do \emph{not} model an end-to-end system and
\atb{2}{they} \emph{cannot} be used to understand the system-wide performance
impacts of DRAM techniques.

\atb{2}{Second}, FPGA-based simulators \om{3}{(e.g.,~\cite{biancolin2019fased,
karandikar2018firesim,mosanu2022pimulator})} model and evaluate modern systems at high simulation
speeds. These platforms have \param{two} drawbacks \mirev{that prevent them from
enabling rapid and accurate DRAM technique evaluation}: (1) \ous{2}{the
simulated system \emph{cannot} operate real DRAM chips because the simulation
interfaces with main memory through a rigid interface} and (2) modifying
FPGA-based simulators requires \mirev{deep} hardware \atb{3}{description}
language (HDL) expertise.

\atb{2}{Third}, FPGA-based \dn{0}{emulators}
\om{3}{(e.g.,~\cite{amid2020chipyard, olgun2022pidram,mosanu2022pimulator,mosanu2023freezetime})} model
end-to-end systems and can \om{3}{use} real DRAM chips.
\ous{2}{However,}
(1) \ous{2}{modifying the memory controller of these systems to implement DRAM
techniques} requires deep HDL expertise and (2) the processor of the
\mirev{modeled} system \mirev{operates at \om{3}{a much lower} clock
\om{3}{frequency (e.g., \SI{50}{\mega\hertz} to \SI{200}{\mega\hertz})}} while
the DRAM operates at a much higher clock \om{3}{frequency} \mirev{(e.g.,
\SI{1.6}{\giga\hertz} \om{3}{to \SI{3.2}{\giga\hertz}})}. This clock frequency
discrepancy limits the accuracy of the evaluation and \om{3}{can} heavily skew
the results.

\begin{table*}[b]
\centering
\caption{Comparison of \X{} with related state-of-the-art prototyping and evaluation platforms.}
\fontsize{7}{7}\selectfont
\begin{tabularx}{\linewidth}{l*{6}{>{\centering\arraybackslash}X}}
\toprule
\textbf{Platforms}
& \textbf{Interface with real DRAM chips}
& \textbf{Flexible \atb{3}{memory controller} for DRAM techniques}
& \textbf{Evaluat\atb{3}{ed CPU clock cycles per second}}
& \textbf{Accurate performance evaluation}
& \textbf{\om{3}{Easily} configurable system}
\\

\midrule
Commercial computing systems & \atb{3}{\cmark} & \xmark & \om{4}{Billions}& \cmark & \xmark \\

Software simulators~\cite{binkert2011gem5,kim2016ramulator,forlin2022sim2pim,power2014gem5,luo2023ramulator2} & \xmark & \cmark (C/C++) & \atb{3}{$\approx{}\!$10K - $\approx{}\!$1M} & \cmark & \cmark \\

\ous{0}{FPGA-based} simulators~\cite{biancolin2019fased, karandikar2018firesim} & \xmark & \xmark & \atb{3}{$\approx{}\!$4M - $\approx{}\!$100M} & \cmark & \cmark \\ 

DRAM testing platforms~\cite{hassan2017softmc,olgun2023drambender} & DDR3/4 & \xmark & \atb{3}{N/A} & \xmark & \xmark \\

\ous{0}{FPGA-based emulators}~\cite{amid2020chipyard, olgun2022pidram, mosanu2022pimulator,mosanu2023freezetime} & DDR3/4 & \ous{0}{HDL} & \atb{3}{50M - 200M} & \xmark & \cmark \\
\midrule
\X{} (this work) & DDR4 & \cmark (C/C++) & \atb{3}{$\approx{}\!$10M} & \cmark & \cmark \\
\bottomrule

\end{tabularx}
\label{tab:tblcompare}
\end{table*}

\atbcrcomment{3}{We discuss PARDIS et al., in related work. I'd argue PARDIS is
not a platform you can immediately use to evaluate DRAM techniques. You could
implement PARDIs in an FPGA, but even a high-perf implementation of PARDIS would
still suffer from issues FPGA emulators suffer from. Programming it would be
easier, though. Perhaps we should say that our programmable memory controller
component builds on such prior work?}Table~\ref{tab:tblcompare} presents a
qualitative analysis of \param{five} \ous{1}{evaluation platform \atb{2}{types}}
that can be used to evaluate DRAM techniques. Among the available \atb{2}{types}
of evaluation platforms, we identify \param{five} main challenges in the design
of an end-to-end evaluation framework for DRAM techniques. The framework needs
to 1) \om{3}{interface with} real DRAM chips and provide correct insights on the
techniques, 2) easily \om{3}{and flexibly} accommodate necessary modifications
(e.g., customize DRAM timings) for DRAM techniques, 3) be fast \om{3}{enough}
\ous{0}{to} allow \atb{2}{rapid} end-to-end full workload evaluation, 4)
\mirev{accurately model the performance of modern computing systems},
and 5) have an easily configurable system to meet user needs (e.g., \atb{2}{to
easily test a wide variety of} processor \atb{2}{designs}).\omcomment{3}{make
better matching with table}

\textbf{Our goal} is to develop a configurable framework that addresses the
outlined challenges, allowing fast and accurate evaluation of emerging DRAM
techniques using real DRAM chips. To this end, we \mirev{design and develop}
\X{}. We observe that DRAM testing platforms
\om{3}{(e.g.,~\cite{hassan2017softmc,olgun2023drambender})} and FPGA-based
emulators \om{3}{(e.g.,~\cite{amid2020chipyard,olgun2022pidram,mosanu2022pimulator,mosanu2023freezetime})} provide
solutions to the \param{first}, \param{third}, and \param{fifth} design
challenges. \atb{2}{We build \X{} on these solutions. To solve the two remaining
challenges and enable rapid end-to-end evaluation of a wide-variety DRAM
techniques, \X{} leverages two key \om{3}{new} ideas.}

First, \X{}\atb{3}{'s system design} \mirev{\atb{3}{comprises} an easily}
programmable core that executes a program to model the memory controller
\ous{1}{(e.g., perform request arbitration and
scheduling).\footnote{\atb{3}{Prior works
(e.g.,~\cite{goossens2013reconfigurable, bojnordi2012pardis}) propose high
performance programmable memory controllers that can dynamically adapt
\om{4}{their command scheduling algorithm, physical address to DRAM address
mapping scheme, power management and periodic refresh techniques} to a workload.
\X{} builds on the basic idea of programmable memory controllers and aims to
provide its users with an easy to program flexible memory controller
substrate.}} \om{3}{We} call such programs \emph{software memory controllers}}.
Compared to a rigid (i.e., \emph{not} programmable) memory controller (typically
implemented using an HDL), \ous{1}{a software memory controller} allows
\mirev{the broader community of expert system designers and architects that lack
deep HDL expertise} to rapidly develop and prototype emerging DRAM techniques.

Second, \X{} advances system state in a way that respects \ous{1}{modern} clock
frequency ratios between different system components (e.g., the processor and
the DRAM interface). \om{3}{In an} FPGA-based prototype where it is difficult to
implement memory controller designs at clock frequencies equaling today's real
processor clock speeds \atb{3}{\X{} faithfully captures the timing behavior of
the modeled system}. To do so, we implement an emulation technique
\ous{1}{\mirev{called} \emph{time scaling}, which} allows \om{3}{each} system
component to be evaluated at a different clock speed than \om{3}{its} FPGA clock
speed (e.g., \mirev{a processor running at \SI{50}{\mega\hertz} FPGA clock
speed interacts with the system as if the processor were clocked at
\SI{4}{\giga\hertz}}), building on the design provided
in~\cite{karandikar2018firesim, biancolin2019fased}. This way, \X{} retains the
benefits (e.g., \atb{3}{interfacing with real DRAM chips and short evaluation time})
of hardware evaluation platforms (e.g., PiDRAM~\cite{olgun2022pidram}) while
overcoming their shortcomings (e.g., inaccurate system performance
\mirev{evaluation}).

\mirev{We validate \atb{3}{\X{}'s time scaling using HDL simulation: (1) an \X{}
implementation with a \SI{100}{\mega\hertz} processor clock frequency using time
scaling to emulate a \SI{1}{\giga\hertz} processor clock frequency and (2) an
\X{} implementation with a \SI{1}{\giga\hertz} (\emph{without} time scaling)
processor clock frequency report execution times that differ by less than 0.1\%
on average across 28 PolyBench~\cite{pouchet2012polybench} workloads.}}
{\atb{3}{We implement a baseline \X{} system that models a} modern NVIDIA Jetson
Nano SoC~\cite{jetsonnano}.} \atb{3}{We show that the \X{} baseline with time
scaling yields a memory latency profile resembling that of the real Jetson Nano
SoC.} \ous{2}{We conduct two case studies to demonstrate \X{}'s extensibility
and \atb{3}{simulation speed}. First, we implement
RowClone~\cite{seshadri2013rowclone} and investigate system performance
improvement of data copy and initialization microbenchmarks with two \X{}
configurations: (1) \emph{\X{} - No Time Scaling} (i.e., 50MHz processor
frequency) and (2) \emph{\X{} - Time Scaling}.
RowClone~\cite{seshadri2013rowclone} with \emph{\X{} - No Time Scaling} and
\emph{\X{} - Time Scaling} improves performance by 306.7$\times{}$ and
15.0$\times{}$ \mirev{on average across all evaluated microbenchmarks},
respectively. \mirev{We conclude from our analysis that FPGA-based
\atb{3}{emulators}} (e.g.,~\cite{olgun2022pidram,amid2020chipyard,mosanu2022pimulator}) that do
\emph{not} faithfully model a modern \mirev{system} skew the results (e.g., by
$\approx{}$20$\times{}$) in favor of DRAM techniques. Second, we implement a
DRAM access latency reduction technique (\mirev{building
on}~\cite{kim2018solar}) and evaluate the end-to-end system performance of
PolyBench\atbcrcomment{3}{Other workloads are difficult to run without proper
system (e.g., Linux) support, which we do not have reliably for EasyDRAM still.}
workloads with \X{} time scaling configuration. Our results show that access
latency reduction improves system performance by 2.75\% \mirev{on average}
across all evaluated workloads.}

We make the following key contributions:
\begin{itemize}
    \item We propose \X{}, an FPGA-based framework \mirev{for end-to-end
    evaluation of DRAM techniques. \X{}} overcomes the limitations of prior \om{3}{works}
    by \mirev{enabling developers to implement DRAM techniques using high-level
    software and faithfully capturing the timing behavior of a modeled modern
    computing system using \om{3}{the} \emph{time scaling} \om{3}{technique}.}
    \item We \atb{2}{develop} the \X{} user-friendly \om{3}{high-level} C++ API
    for designing and extending \mirev{\X{}-based} \mirev{software} memory
    controllers. \atb{3}{The high-level API makes} DRAM \om{3}{techniques}
    \mirev{easier to implement} and \mirev{their evaluation studies} more
    accessible to system designers who lack \mirev{deep hardware description
    language} expertise.
    \item We perform two case studies using \X{} and \mirev{demonstrate} that
    state-of-the-art \om{3}{real-system-based} evaluation methodologies
    \mirev{that} do \emph{not} faithfully model a real system can \mirev{yield
    \om{3}{largely}} inaccurate \mirev{system performance results} \om{3}{even
    though they use real hardware}.
    \item \mirev{To enable reproducibility and aid future research in a
    transparent manner, we open-source our \X{} implementation at
    \url{https://github.com/CMU-SAFARI/EasyDRAM}.}
\end{itemize}
\section{Background}

We provide a brief background on DRAM organization, DRAM timing parameters, and
memory controllers. For more detailed background on these, we refer the reader
to many prior
works~\cite{carter1999impulse,chang2016understanding,chang2014improving,
chang2016lisa,chang2017understanding,frigo2020trrespass,ghose2018what,
hassan2016chargecache,hassan2019crow,hassan2017softmc,olgun2023drambender,ipek2008self,
olgun2024abacus, canpolat2024breakhammer}.

\subsection{DRAM Organization}

\begin{figure}[!h]
\centering
\includegraphics[width=\columnwidth]{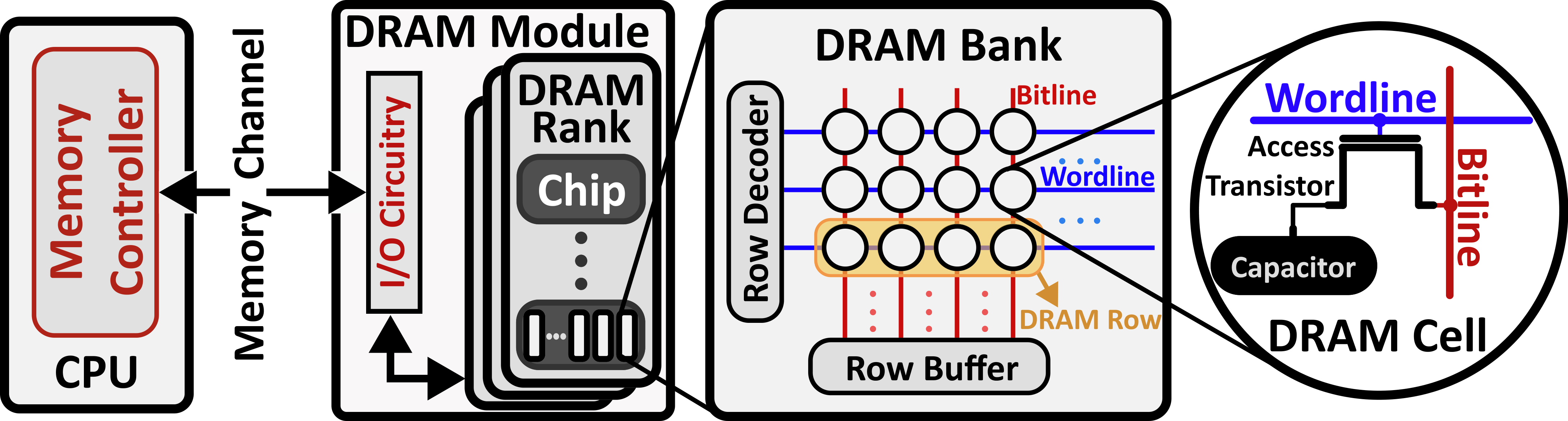}
\caption{DRAM \atb{5}{channel, module, rank, and bank} organization}
\label{fig:dramhierarchy}
\end{figure}

Figure~\ref{fig:dramhierarchy} illustrates the hierarchical structure of
DRAM-based main memory. Typically, a system's \textit{memory controller} links
to several DRAM \textit{modules} across multiple DRAM \textit{channels}, each
operating independently. Each DRAM \textit{module} encompasses one or more
ranks, with each rank composed of several DRAM \textit{chips} functioning in
\atb{4}{lockstep}. 
The memory controller can interact with multiple DRAM ranks by alternating
the channel's I/O bus among the ranks. 

Each DRAM chip is partitioned into multiple \textit{banks}. These DRAM banks
within a chip share common I/O circuitry. The cells within a DRAM bank are
structured into a \om{4}{two-dimensional array} of \textit{rows} and
\textit{columns}. Each \textit{cell} within a DRAM row connects to a common
\textit{wordline} via \textit{access} transistors. A \textit{bitline} links a
column of DRAM cells to a sense amplifier, facilitating data access and
\atb{4}{update}.


\subsection{DRAM Timing Parameters}

The memory controller accesses DRAM locations via three primary steps. First,
the memory controller sends an activate (\textit{ACT}) command to \ous{2}{open}
a closed row within a bank, preparing the row for column access. Second, the
memory controller issues a read (\textit{RD}) or write (\textit{WR}) command to
read from or write to a column in the row, respectively. Third, once all column
operations on the \ous{2}{open} row are completed, the memory controller issues
a precharge (\textit{PRE}) command that closes the row and readies the bank for
a new activation.

DRAM cells are inherently leaky and lose their charge over time due to charge
leakage in the access transistor and the storage capacitor. To maintain
data integrity, the memory controller periodically refreshes each row in a time
interval called \emph{the refresh window} ($t_{REFW}$) ($32 ms$ for
DDR5~\cite{jedec2020ddr5} and $64 ms$ for DDR4~\cite{jedec2017ddr4} \om{4}{at
normal operating temperature range of \SI{0}{\celsius} -- \SI{85}{\celsius}}).
\atbcrcomment{5}{0-85 is normal for at least a class of chips not deisgned to withstand -40 degrees celsius}
To refresh all rows every $t_{REFW}$, the memory controller issues
REF commands with a time interval called \emph{the refresh interval} $t_{REFI}$
($3.9 \mu s$ for DDR5~\cite{jedec2020ddr5} and $7.8 \mu s$ for
DDR4~\cite{jedec2017ddr4}).

To ensure correct DRAM operation, the memory controller must adhere to
standardized timings between consecutive commands, known as \textit{timing
parameters}~\cite{kim2012case,lee2015adaptive,lee2013tiered}. These parameters
ensure the internal DRAM circuitry has adequate time to execute the operations
required by the command~\cite{orosa2021codic}.

\subsection{Memory Controller}

A memory
controller \ous{2}{\atb{4}{orchestrates} data \atb{4}{transfers}} between the
\atb{4}{processor and DRAM}. It plays a role in ensuring efficient memory
utilization, \om{4}{improving} system performance \atb{4}{and energy
consumption}, \atb{4}{enabling fair memory access,} and maintaining \atb{4}{DRAM
data integrity~\cite{zuravleff1997frfcfs, rixner2000memory, parbs-isca08,
stfm-micro07,subramanian2014bliss,subramanian2016bliss, carter1999impulse,
ipek2008self}}. The memory controller's responsibilities include \om{4}{the
following.}

\noindent\textbf{Arbitration of Requests.} The memory controller arbitrates
memory requests from various sources \atb{4}{(e.g., hardware threads) in a
system}. The arbitration is critical to \om{5}{minimize}
contention and ensure fair and efficient access to shared memory resources.

\noindent\textbf{Physical to DRAM Address Translation.} The memory controller
\atb{4}{translates} physical memory addresses (e.g., provided by the
\dn{0}{\atb{4}{memory management unit} in the \atb{4}{processor}}) into DRAM
channel, rank, bank, row, and column addresses.

\noindent\textbf{Policy-Based Request Servicing.} The memory controller employs
scheduling algorithms to determine the order in which the buffered requests are
serviced. \atb{4}{Each scheduling algorithm,} such as First Ready, First Come,
First Serve (FR-FCFS)~\cite{zuravleff1997frfcfs, rixner2000memory}, \om{4}{or
PAR-BS~\cite{parbs-isca08}}, aim\atb{4}{s to achieve a different goal: some
maximize memory throughput (e.g., FR-FCFS), others improve fairness across
memory request sources while maximizing \om{5}{system performance} (e.g.,
PAR-BS).} The memory controller must consider the state of the DRAM
\atb{4}{(e.g., a row is open)}, timing constraints, and the type of requests
\atb{4}{(e.g., read or write)} for scheduling.

\noindent\textbf{Issuing DRAM Commands.} The memory controller determines
\om{5}{a correct and high-performance} sequence of DRAM commands to
serve requests and issues \atb{4}{DRAM commands} through the \atb{4}{DRAM
physical interface (e.g., DDR4 or DDR5)}.

The interaction of these functions within a memory controller is complex, with
each aspect impacting system performance. Therefore, the design and
implementation of memory controllers require \ous{0}{carefully considering}
these elements~\cite{subramanian2014bliss,subramanian2016bliss} to
\ous{0}{balance} system speed, energy \atb{4}{consumption}, fair memory
access\om{5}{, and hardware complexity}.
\section{Motivation}
\label{sec:motivation}

\mirev{Evaluating the system-wide benefits of DRAM techniques is non-trivial as
they require modifications \mirev{across} multiple layers of the
\mirev{\atb{4}{computing}} stack (as described in~\secref{sec:introduction}).
Table~\ref{tab:tblcompare} compares existing platforms that could be used to
evaluate system integration of DRAM techniques.} We find FPGA-based
\mirev{platforms} to be a good \atb{4}{basis for} \atb{4}{designing a flexible
and easy-to-use DRAM technique evaluation platform due to} \param{three}
reasons. First, \atb{4}{an FPGA-based platform can easily interface with} real
DRAM chips \atb{4}{and expose a low-level DDRx interface to its user, enabling
the user to \om{5}{gain} reliable insight into real system integrations of DRAM
techniques}. Second, \atb{4}{an FPGA-based model of a real system is fast
enough for rapid full workload evaluation.} Third, existing
open-source system design libraries~\cite{amid2020chipyard,openhwwebsite} allow
users to easily configure the modeled system.

However, \ous{1}{an FPGA-based \atb{4}{platform is challenging to directly use
for DRAM technique evaluation because of two main reasons.} First, an FPGA
design is hard to model and extend \mirev{because \atb{4}{doing so} requires
deep} hardware design expertise. Second, an FPGA design \emph{cannot} accurately
accommodate a modern system with high design complexity. For example, \mirev{an
FPGA implementation of a} processor \atb{4}{is typically clocked at
\SI{50}{\mega\hertz} --
\SI{200}{\mega\hertz}}~\cite{asanovic2016rocket,openhwwebsite} (\om{5}{1-2}
orders of magnitude slower than a \om{5}{real}
processor~\cite{berry2020ibm,geva2022ibm})}, while the DRAM \atb{4}{interface is
clocked at a much higher clock frequency (}e.g., \SI{1.6}{\giga\hertz} --
\SI{3.2}{\giga\hertz}~\cite{jedec2017ddr4}).

\begin{figure}[!h]
\centering
\includegraphics[width=\columnwidth]{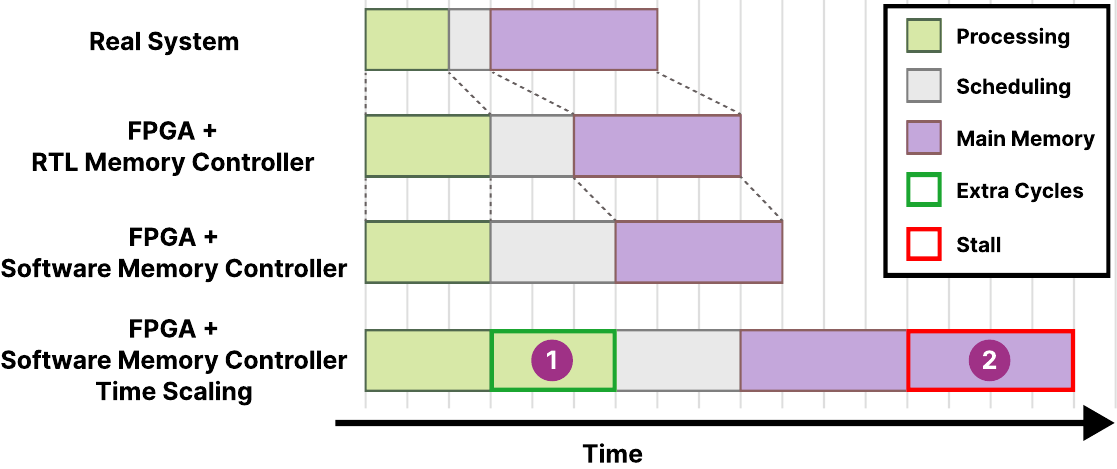}
\caption{\ous{2}{Execution \atb{4}{time breakdown \atb{5}{(plotted qualitatively
for motivation purposes)} of a memory request}}}
\label{fig:timescalemotivation}
\end{figure}

\atb{4}{To pictorially explain the second challenge,}
\figref{fig:timescalemotivation} depicts \atb{4}{the execution time breakdown 
\atb{5}{(plotted qualitatively for motivation purposes)}
of a memory request into time spent by 1)~the processor, 2)~the memory scheduling
algorithm, and 3)~main memory.} 
\atb{4}{We show the breakdown for four system configurations: 1)~a real system,
and an FPGA implementation of the real system with 2)~a hardware (RTL) memory
controller, 3)~a software memory controller, and 4)~a software memory
controller augmented with the \emph{time scaling} technique (introduced
in~\secref{sec:introduction}).}

We make \param{three} observations from \figref{fig:timescalemotivation}. First,
FPGA \atb{4}{implementations of the real system} take longer to execute the same
\atb{0}{number} of processor and scheduling cycles compared to a real system.
Second, \atb{4}{a} software memory controller 
takes more time \atb{4}{to schedule a memory request} than \atb{4}{an}
\dn{0}{RTL} memory controller. Third, although \atb{4}{an FPGA implementation of
the real system takes longer to execute a memory request,} the DRAM access
latency stays constant \atb{4}{(that is, the Main Memory bar stays the same
length across the first three systems)} as \atb{4}{DRAM performs the same across
all of these systems (i.e., has identical operating
frequency and timing constraints).}

Based on these observations, we \atb{4}{conclude} that an FPGA-based
platform\atb{4}{, when used for end-to-end evaluation of DRAM techniques,} is
\atb{4}{likely to yield inaccurate system performance results because the
FPGA-based platform} combines \atb{4}{a relatively high-clock-frequency} real
DRAM chip with \atb{4}{a low-clock-frequency} processor. \atb{4}{To enable
\emph{accurate} and rapid evaluation of DRAM techniques end-to-end, easily
\emph{without} needing deep hardware design expertise, we design 
EasyDRAM.}

\section{\X{}}
\label{sec:mechanism}

End-to-end evaluation of DRAM techniques \atb{0}{using} real DRAM chips requires
modifications to multiple levels of the comput\om{5}{ing} stack (e.g., \atb{0}{memory
controllers, memory mapping and allocation in the operating system, and user
applications}). To ease the evaluation process, we propose \X{}, a framework
that allows rapid implementation and accurate end-to-end evaluation of DRAM
techniques with real DRAM chips.

\begin{figure}[h]
\centering
\includegraphics[width=0.90\linewidth]{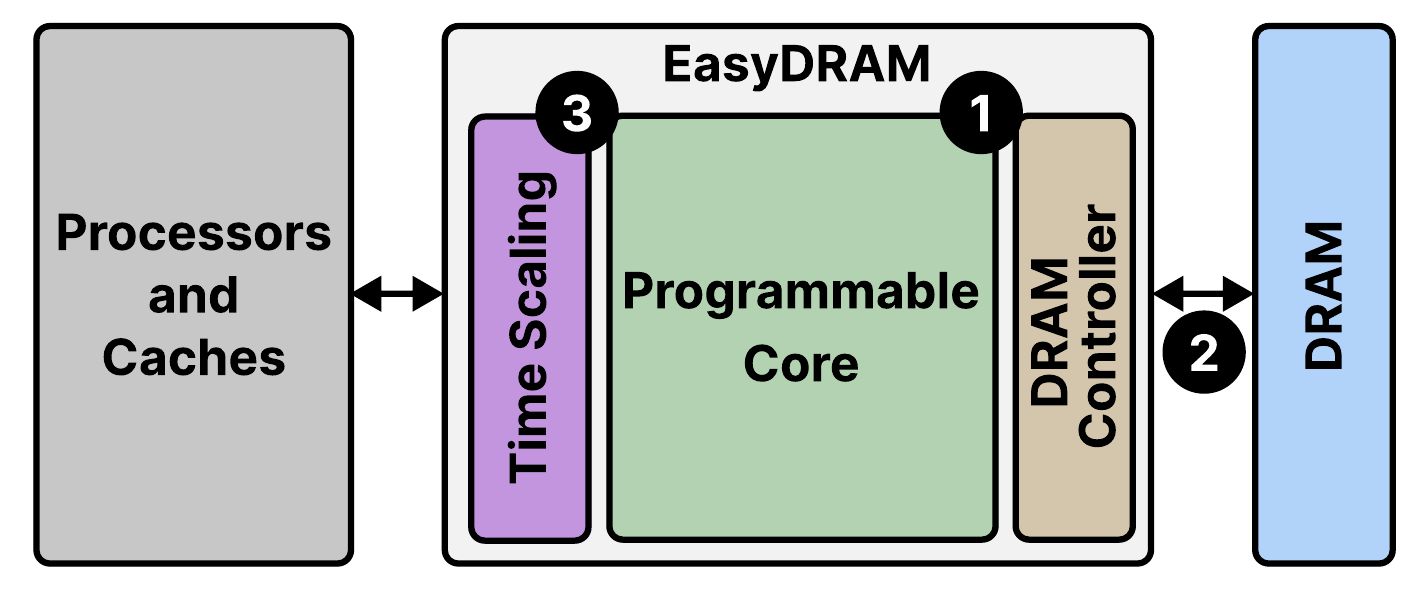}
\caption{\mirev{\X{} high-level overview}}
\label{fig:easyoverview}
\end{figure}

\head{\mirev{Overview}} We design \X{} to 1) be easy to use, 2) allow evaluation
on real DRAM chips, and 3) provide \emph{accurate} system performance results.
\mirev{\figref{fig:easyoverview} depicts a high-level overview of \X{} in a
simplified computing system.}

\mirev{\X{} is implemented between the last-level cache and the main memory.}
\X{} has \param{three} key ideas that enable our design goals.

\head{\circled{1}~\mirev{Programmable Core}} \X{} implements a programmable core
as a memory controller that allows easy modifications with software (i.e., C++)
changes without requiring hardware description language (HDL, e.g., Verilog)
expertise.

\head{\circled{2}~\mirev{Operating Real DRAM Chips}} \X{} interfaces with real
DRAM chips and provides an easy-to-use API to the user that allows issuing
arbitrary DRAM commands with arbitrary delays between any two consecutive
commands.

\head{\circled{3}~\mirev{Time Scaling}} \X{} overcomes the accurate performance
evaluation limitations of FPGA-based systems with \emph{time scaling}. Time
scaling 1) separates hardware components (e.g., memory controller) to different
emulation domains and 2) provides fine-grained control over the emulation
duration and \ous{2}{emulat\atb{5}{ed}} \atb{5}{clock frequency} of each domain.
By doing so, \X{} emulates each hardware component to match \om{5}{expectations
from a realistic processor model}.

\subsection{Programmable Memory Controller}
\label{sub:progcontroller}

Implementing new DRAM techniques requires invasive modifications to memory
controller design. Existing frameworks do \emph{not} provide the means to easily
make these changes. Extending FPGA-based
emulators~\cite{amid2020chipyard,olgun2022pidram} to evaluate DRAM techniques
\om{5}{is} time consuming for three reasons. First, extending FPGA-based
emulators require\om{5}{s} HDL and \atb{5}{hardware design}
expertise. Second, FPGA-based emulators need to go through lengthy synthesis,
floor-planning, placement, and routing steps even for the smallest changes.
Third, FPGA-based emulators require components to meet \emph{timing
constraints}. As such, evaluating DRAM techniques using FPGA-based emulators
\om{5}{(and thus HDLs)} ha\om{5}{s} relatively high overheads at multiple stages
(e.g., design, debugging, emulation) of the evaluation process. 

To overcome these challenges, \X{} \atb{0}{implements and uses} a \emph{fully
programmable \atb{0}{general-purpose processor}} instead of \atb{0}{a}
conventional RTL memory controller. This way, memory scheduling decisions or
DRAM operations are expressed in software without extensive HDL expertise, and
results are observed by operating \emph{real} DRAM chips.

Listing~\ref{lst:progctrl} presents a \atb{0}{simple memory controller modeled
using C++ in \X{}}. In this example, the \atb{0}{memory controller} checks for incoming requests
(line 3), operates the DRAM chip (lines 6-8), and responds with the obtained
data (lines 9-10). \X{} provides an easy means to perform low-level operations
such as buffering the requests and executing the DRAM commands.

\begin{lstlisting}[float, floatplacement=h, language=C++, caption={\mirev{A \om{5}{C++} memory controller 
for EasyDRAM \om{5}{serving only read requests}.}}, label=lst:progctrl]
void cpp_controller() {
    // Wait for a request to arrive
    while (req_empty()); 
    // Move the request from buffer to scratchpad 
    Request req = receive_request(); 
    // Translate physical address to  DRAM address
    auto addr = get_addr_mapping(req.addr); 
    // Issue DRAM commands to serve the request
    read_sequence(addr);
    flush_commands(); 
    rdback_cacheline(req.data); 
    // Send request response to the processor
    enqueue_response(req); 
}
\end{lstlisting}

\head{Challenges of Implementing a Programmable Memory Controller}
Using a programmable core \atb{5}{to execute a software memory controller}
within an FPGA-based system has \param{two} challenges. First, DRAM operation
requires obeying strict timing parameters. \atb{5}{The software memory
controller} \emph{cannot} issue DRAM commands at nanosecond granularity.
\ous{0}{The software memory controller \atb{5}{can only} issue a DRAM command
\atb{5}{after it processes} a memory request.} The \atb{5}{memory controller
(e.g., the C++ program in Listing~\ref{lst:progctrl}) executes hundreds of
instructions in the} programmable core to process a memory request
\atb{0}{and issue a DRAM command}, which may take hundreds of \atb{5}{FPGA}
clock cycles. Second, a software memory controller can
receive a request every few \atb{5}{FPGA} clock cycles for memory-intensive
workloads. \atb{5}{Therefore,} the programmable core's request queue remains
full for a long time (because the \ous{0}{software memory controller} can take
hundreds of \ous{0}{\atb{5}{FPGA clock}} cycles to process a memory request).
\ous{1}{As such}, the processor \emph{cannot} execute memory instructions for
hundreds of cycles. \ous{1}{\atb{5}{The processor in the FPGA-based system with
a software memory controller experiences lengthier stalls than in a real system.
This discrepancy reduces the evaluation accuracy of the FPGA-based system}.}

\X{} addresses these two challenges. First, \X{} \atb{5}{decouples the interface
between the software memory controller and DRAM. The software memory controller
(relatively slowly) prepares a list of DRAM commands and command timings needed
to serve a single or a series of memory requests. The software memory controller
provides the list of DRAM commands and timings through the decoupled interface
to logic specialized for executing that command sequence at nanosecond
granularity. To implement the specialized logic, we leverage design reuse
principles and heavily reuse the open source DRAM testing platform, DRAM
Bender~\cite{olgun2023drambender,safari-drambender}}. Second, \X{} uses
\emph{time scaling}, which decouples the \atb{5}{relatively slow software memory
controller} from \atb{5}{relatively fast} system components that generate memory
requests (e.g., the \ous{0}{processor}) \atb{5}{and carefully advances the system
state} \ous{0}{to \atb{5}{faithfully capture the timing behavior of a modeled
system}}. The following section provides a deeper look into the challenges and
explains \atb{5}{how we} overcome \ous{0}{the challenges}.

\subsection{Operating Real DRAM Chips}
\label{sub:realdram}

DRAM operation requires timely execution of DRAM commands. Modern systems
include fast memory controllers that issue commands to the DDRx interface.
Performing these actions with \om{5}{a} software memory controller is
\ous{0}{challenging for \param{two} reasons. First, \om{5}{a} software memory
controller executes hundreds of instructions to produce a DRAM command.
Therefore, \om{5}{a} software \atb{5}{memory controller} \emph{cannot} quickly
(i.e., every few nanoseconds) decide on commands to be issued. Second,
general-purpose programmable cores (e.g., RISC-V cores) on FPGAs have clock
frequencies of \atb{5}{\SI{50}{\mega\hertz} --
\SI{200}{\mega\hertz}}~\cite{asanovic2016rocket,openhwwebsite}. Therefore, a
\atb{5}{software memory} controller \emph{cannot} reliably operate \ous{2}{the
high-speed DDRx interface (e.g., \atb{5}{\emph{cannot} issue} a command every
\atb{5}{few hundred picoseconds or every few
nanoseconds~\cite{jedec2017ddr4}})}.} 

\begin{figure}[h]
\centering
\includegraphics[width=\columnwidth]{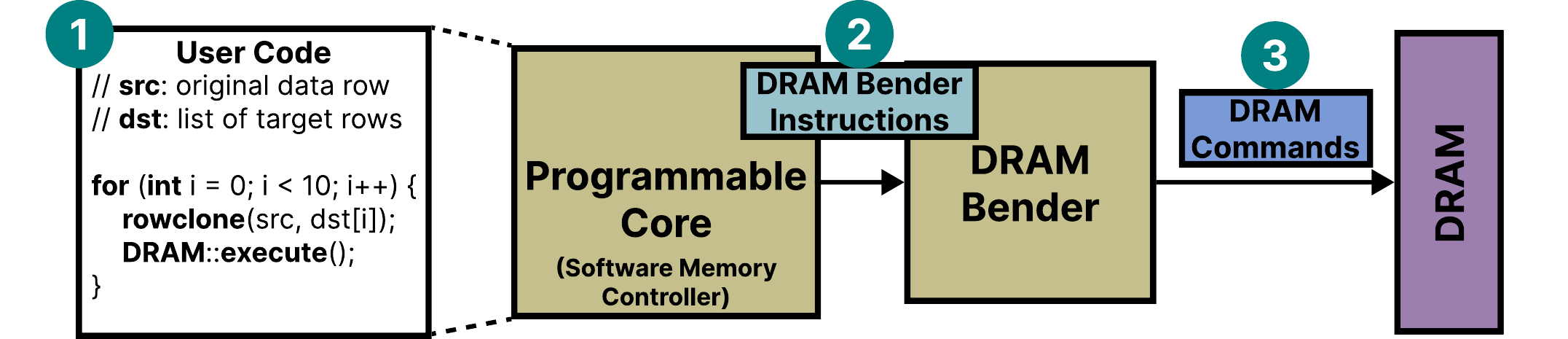}
\caption{\ous{0}{High-level \atb{5}{p}rogrammable core and DRAM Bender execution
model}}
\label{fig:easydrambender}
\end{figure}

\X{} overcomes these challenges by \atb{5}{interfacing the software memory
controller} with DRAM Bender~\cite{olgun2023drambender,safari-drambender}, the
state-of-the-art DRAM testing platform. DRAM Bender allows users to issue DRAM
commands using the DRAM Bender instruction set architecture (ISA). \X{}
introduces a command buffer shared between the \atb{5}{software memory
controller} and DRAM Bender. The software memory controller fills this buffer
with DRAM Bender instructions to \atb{0}{issue DRAM commands}. \atb{0}{As an
example}, \figref{fig:easydrambender} presents \ous{0}{how an in-DRAM copy
technique (i.e., RowClone~\cite{seshadri2013rowclone})} can be
\ous{0}{implemented} in a simplified \atb{0}{depiction} of this system.

\atb{0}{The} user implements a program \circledteal{1} \ous{0}{that copies} the
content of a row to multiple target rows. The \ous{0}{software} memory
controller \ous{0}{running on the programmable core} offloads the DRAM commands
and their timings to DRAM Bender \dn{0}{as DRAM Bender instructions}
\circledteal{2} and the \emph{execute} call triggers DRAM Bender to start
issuing \ous{0}{commands to DRAM} \circledteal{3}.

\ous{0}{The software memory controller generates the DRAM commands and waits for
DRAM Bender to execute the commands. Therefore}, \X{}'s DRAM operation
\ous{0}{contains} \param{two} types of time-slices: \ous{0}{1) DRAM is idle
(i.e., generating the DRAM commands) and 2) DRAM is operated in real-time (i.e.,
DRAM Bender execution).} The high command scheduling latency of the software
memory controller and overheads of being coupled with DRAM Bender (e.g.,
transferring DRAM \ous{2}{commands}) need to be considered for realistic system
evaluation. To do so, \om{5}{we develop} \emph{time scaling} \om{5}{and
implement it in \X{}}.

\subsection{Time Scaling}
\label{sub:timescale}

\om{5}{A} software memory controller runs significantly slower than a hardware
memory controller. We identify two problems with the evaluation accuracy of a
slow memory controller. First, serving a memory request slower causes additional
memory requests to arrive and potentially change subsequent scheduling decisions
(e.g., keeping a DRAM row open). Second, a processor can execute additional
instructions due to increased memory access latency. Both of these problems
reduce evaluation accuracy because the executed instruction traces and issued
DRAM commands as \om{5}{a} result of differing scheduling decisions do \emph{not}
faithfully emulate a real system.

We \atb{5}{devise a mechanism that allows
a software memory controller to identify and ignore memory requests
that arrive due to the software memory controller's slowness. This mechanism
solves the first problem with the evaluation accuracy of a slow memory controller.}
For the second problem, we divide \atb{5}{the modeled} system into
\atb{5}{multiple} emulation domains and \atb{5}{enable} fine-grained control of the progression of
each domain. By doing so, \X{} can stop a processor from executing additional
instructions due to the increased memory access latency of a software memory
controller.

To control the emulation \atb{5}{clock frequency} of different hardware
components, we \atb{5}{develop} a mechanism called \emph{time scaling}. Time
scaling \ous{2}{aligns} the emulation rate of each domain to the slowest
progressing domain (i.e., the domain with the most discrepancy between its real
and FPGA clock frequency). To do so, time scaling \ous{2}{stalls} \atb{5}{a
faster progressing domain (e.g., the processor)} thereby providing \atb{5}{the}
slower domain extra cycles to satisfy the desired system clock frequency
requirement. \X{} implements \emph{time scaling counters} to track the emulation
point (i.e., emulated clock cycle) for each emulation domain (e.g., \atb{5}{the
processor}). By doing so, \X{} accurately tracks each emulated domain's
emulation \atb{5}{progress}.

\begin{figure*}[!th]
\centering
\includegraphics[width=0.9\linewidth]{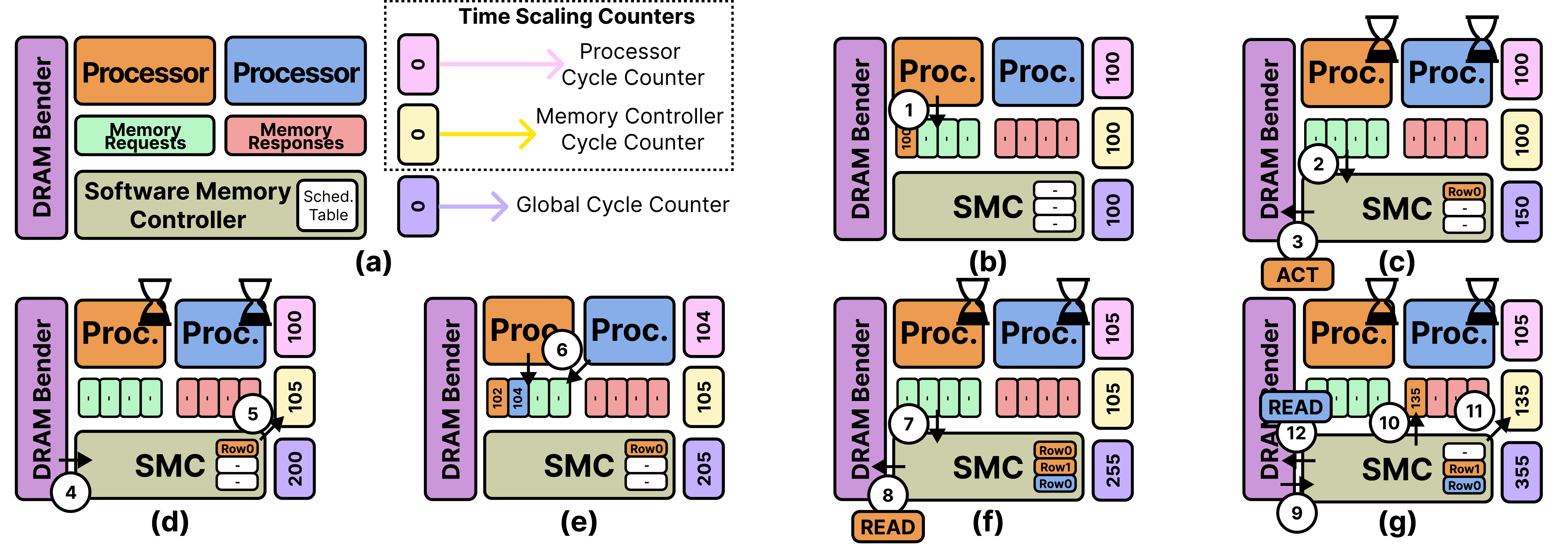}
\caption{Time scaled execution of \X{} at different \ous{0}{emulation} steps}
\label{fig:easyexecution}
\end{figure*}

\atbcrcomment{5}{I don't have a quick solution for too many things happening
in (g) but I heavily revised the text to hopefully make it more clear.}
\figref{fig:easyexecution} visualizes \ous{2}{\X{}'s state} at various stages of
the emulation process. \atb{5}{\figref{fig:easyexecution}-(a)-left shows a
simplified diagram of an example \X{} system with two processors (top), a
software memory controller (bottom), and memory request and response queues
(middle).} \atb{5}{\figref{fig:easyexecution}-(a)-right shows \emph{time scaling
counters}.} The time scaling counters for the \atb{5}{processors}
\atb{5}{(processor cycle counter)} and a software memory controller
(\atb{5}{memory controller cycle counter}) are used to track the progress of
their respective emulation points. The global counter serves as a reference
timer and counts clock cycles from the system power-on \om{5}{time}. We note
that all \atb{5}{processors} share the same time scaling counter, \ous{1}{and}
synchronization actions \atb{5}{taken by \X{} to stall an emulation domain}
(e.g., clock-gating) affect all processors. As the emulation starts, all
counters are initialized to 0 and no \atb{5}{memory} requests or responses
exist. Time scaling counters are incremented every cycle while the system has no
unresolved memory requests.

As the emulation proceeds (b), \atb{5}{processor (left)} makes the first main
memory request~\circledmt{1}. The request is tagged with the current
\atb{5}{processor cycle} counter value \atb{5}{(100)} and the
\atb{5}{processors} get clock-gated \atb{5}{(indicated by an hourglass)},
stopping their execution until the memory request is processed.
When \atb{5}{software memory controller (SMC)} detects the request (c), it
enters \emph{critical mode}. In critical mode, SMC locks the processor
\atb{5}{cycle} counter such that the \atb{5}{processor}
\emph{cannot} emulate ahead of SMC. By doing so, 
SMC controls the processor emulation rate to match the targeted clock frequency
constraints. Then, SMC transfers the incoming request from hardware buffers to
software memory~\circledmt{2}. SMC makes a scheduling decision, creates a row
activate (ACT) command, and sends the command to DRAM Bender~\circledmt{3}.
Once DRAM Bender finishes execution (d), SMC obtains the time spent for the ACT
command \om{5}{from DRAM Bender}~\circledmt{4}, calculates the
cycles the processor domain should execute at the emulated system's clock
frequency, and advances the \atb{5}{memory controller cycle counter} by
\atb{5}{the number of cycles spent for the ACT command}~\circledmt{5}.
Updating the time scaling counter \atb{5}{enables} the processors \atb{5}{(i.e.,
the processors are no longer stalled by clock-gating)} and the processors
emulate the missing time scaled duration (e) where processors send \param{one}
new request \atb{5}{each}~\circledmt{6}.
When the processor \atb{5}{cycle} counter reaches \atb{5}{the value of
the memory controller cycle} counter (f), SMC transfers new requests from
hardware buffers to software memory~\circledmt{7}.\footnote{\atb{5}{By doing so,
\X{} ensures that a relatively slow software memory controller can observe all
memory request created by the processors before making the next scheduling
decision.}} Then, SMC makes the next scheduling decision and issues a read
command~\circledmt{8}.

After executing the read command (g), SMC obtains the data and time spent for
execution~\circledmt{9}. SMC tags the response with \atb{5}{a value (in terms of
processor cycle count) that indicates when a processor is allowed to consume
that response} and writes \atb{5}{the response} to the hardware response
buffer~\circledmt{10}. \atb{5}{Doing so ensures that the processor does
\emph{not} observe a memory request response earlier than expected and execute
additional instructions that depend on this response ahead of time.} The
duration spent on scheduling a \atb{5}{memory request} is converted to the number
of emulation cycles at the emulated system's clock frequency and the
\atb{5}{memory controller cycle} counter is updated~\circledmt{11}. 
The processors do \emph{not} send new
requests and SMC issues another read command~\circledmt{12} to the open row.
\atbcrcomment{5}{step 12 can be removed for a less crowded figure}

The system runs similarly to our example as long as there are requests in the
hardware buffer or SMC is in critical mode (e.g., SMC has \emph{not} \atb{5}{yet
processed and responded to all memory requests}). When critical mode ends, the
time scaling counters synchronize \atb{5}{as processors are \emph{not}
clock-gated (the processor cycle counter is incremented each global clock cycle
to match the memory controller cycle counter)} and processors continue their
normal execution.

\subsection{Lifetime \om{5}{of a Memory} Request}
\label{sub:lifetime}

\begin{figure}[h!]
\centering
\includegraphics[width=\columnwidth]{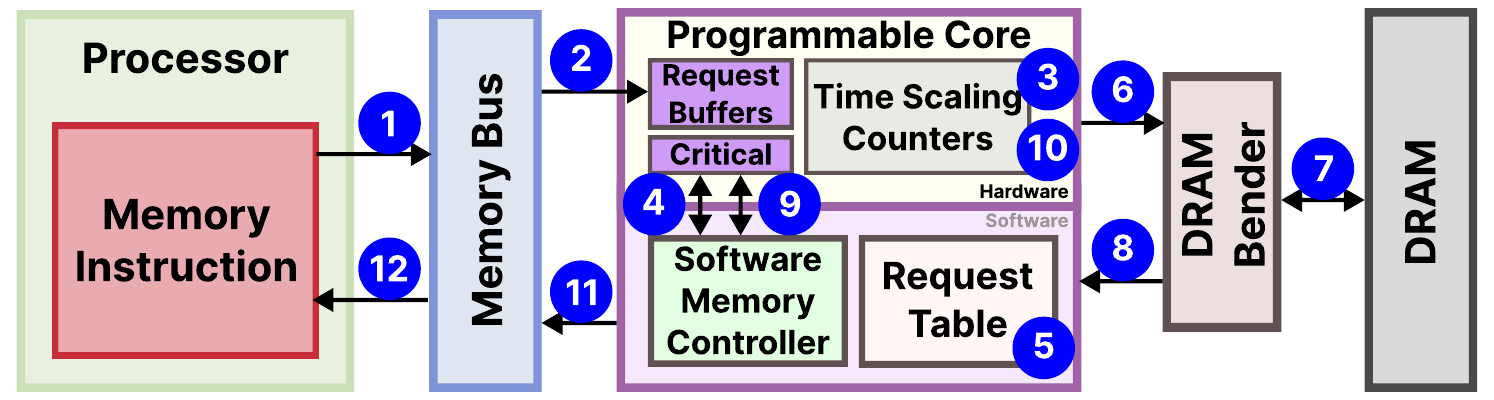}
\caption{Lifetime of a memory request with \X{}}
\label{fig:lifetimeofreq}
\end{figure}

\omcomment{5}{Maybe show earlier}
Figure~\ref{fig:lifetimeofreq} \ous{1}{outlines the general execution of a
memory request.}

The \dn{1}{processor} initiates a \atb{5}{memory} read request~\circledblue{1}.
This request is placed into \atb{5}{the programmable core's} hardware
buffers~\circledblue{2} and the processor \ous{0}{\atb{5}{gets
clock-gated}}~\circledblue{3}. The software memory controller (SMC) implements a
loop that i) checks for new requests, ii) makes a scheduling decision, and, iii)
handles responses from DRAM. When \ous{2}{a request is found, SMC} enters
critical mode~\circledblue{4} and transfers the new request from the hardware
buffers to the software request table~\circledblue{5}. \ous{2}{SMC makes a
scheduling decision and transfers the} DRAM commands to DRAM
Bender~\circledblue{6}. DRAM Bender \ous{2}{executes the DRAM
commands~\circledblue{7}, then} returns the data and the number of cycles
\ous{2}{taken for execution}~\circledblue{8}. SMC \ous{2}{finalizes the memory
request response, \atb{5}{tags the response with the processor cycle count value
indicating when the response can be consumed by the processor,} and} transfers
\ous{2}{the response} to the \ous{2}{request} buffers~\circledblue{9}.
\ous{2}{Lastly, SMC \atb{5}{advances} the \atb{5}{memory controller time
scaling} counter~\circledblue{10} and exits critical mode}. As the
\ous{2}{execution continues}, the processor time scaling counter reaches
\ous{2}{the response's} \atb{5}{tagged counter value}. The response is
transferred to the memory bus~\circledblue{11}, completing the request's
lifetime by reaching the processor~\circledblue{12}.
\setcounter{version}{6}

\section{\X{} Implementation}
\label{sec:implementation}

Our \X{} implementation takes advantage of agile hardware development platforms,
\om{6}{particularly,} the Chipyard Framework~\cite{amid2020chipyard}. Chipyard
facilitates the design and evaluation of full-system hardware in an agile
environment. We use Chisel~\cite{bachrach2012chisel} as our primary HDL.

\head{System Implementation}
We \om{6}{use} Chipyard for two key benefits. First, Chipyard contains a diverse
selection of processors and offers high system configurability. This variety
provides a strong base for modifications and experimentation. Second, updates to
Chipyard enhance the system feature availability, ensuring that our \X{}
implementation has access to future advancements. \mirev{Our default \X{} system
implements the Berkeley Out-of-Order RISC-V Processor
(BOOM)~\cite{celio2018thesis} as the processor.}

\head{Programmable Controller}
We use Rocket~\cite{asanovic2016rocket} as our programmable core. We modify the
Rocket Core by separating the core from the \emph{RocketTile} and disabling
virtual memory support. Then, we connect its memory ports to EasyTile's
interconnect.

\begin{table*}[!b]
\centering
\caption{Commonly used hardware \atb{6}{abstraction} and software library
functions}
\vspace{-1em}
\fontsize{10}{8}\selectfont
\begin{center}
\renewcommand{\arraystretch}{1.5}
\begin{tabular}{|l|l|} 
\hline
\textbf{Function Name} & \textbf{Description} \\ 
\hline
\hline
\emph{set\_scheduling\_state(bool state)} & \ous{2}{Set critical mode register}
\\ 
\emph{get\_request()} & \ous{2}{Read a request from the hardware request buffer}
\\ 
\emph{ddr\_activate()/precharge()/read()...} & \ous{2}{Insert a DRAM command to
the command batch} \\ 
\emph{flush\_commands()} & \ous{2}{Execute the DRAM commands in the command
batch} \\
\hline
\hline
\emph{add\_request(Request\& req)} & \ous{2}{Insert a \atb{6}{memory} request to
the request table in \atb{6}{programmable core memory}} \\ 
\emph{FRFCFS::schedule()} & \ous{2}{Select a request with the FR-FCFS scheduler}
\\ 
\emph{FCFS::schedule()} & \ous{2}{Select a request with the FCFS scheduler} \\ 
\emph{rowclone(Address src, Address dst...)} & \ous{2}{Insert a RowClone command
sequence to the command batch} \\ 
\hline
\end{tabular}

\label{tbl:apigeneral}
\end{center}
\end{table*}

\subsection{\X{}'s Hardware Design (EasyTile): Overview}
\label{sub:overview}

\begin{figure}[h]
\begin{center}
\includegraphics[width=1\linewidth]{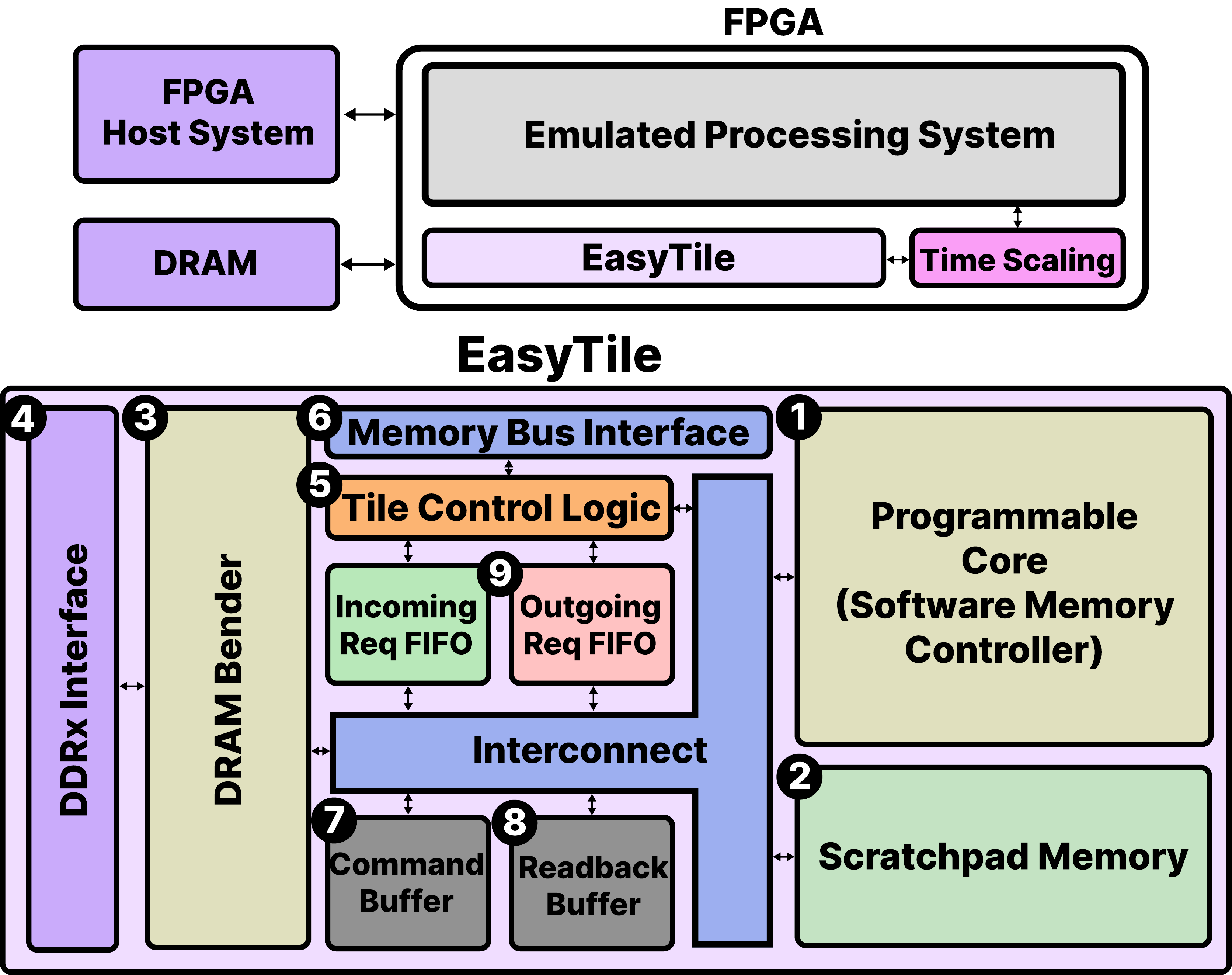}
\end{center}
\caption{The system overview of \X{}}
\label{fig:easydramsystem}
\end{figure}

Figure~\ref{fig:easydramsystem} shows an overview of the main components of \X{}
and its configuration for a\atb{6}{n emulated processing system}.

We pack the programmable core and DRAM Bender with various helper components of
\X{} inside a module we call EasyTile (Figure~\ref{fig:easydramsystem},
\atb{6}{bottom}).
\ous{2}{EasyTile includes the following hardware components.}

\head{Programmable Core~\circled{1}} The programmable core executes \ous{2}{a
program that implements a software memory controller}. \ous{2}{These programs
use the on-tile \atb{6}{scratchpad} memory~\circled{2} to store instructions and
data}. Section~\ref{sec:easyapi} describes the proposed API \ous{2}{that
simplifies programming a} software memory controller.

\head{DRAM Bender~\circled{3}} DRAM Bender allows users to issue DRAM commands
without dealing with the low-level DRAM interface (e.g., DDRx
interface~\circled{4}). Programs running on the programmable core use DRAM
Bender's interface to implement DRAM operations and techniques.

\head{Tile Control Logic~\circled{5}} Tile Control Logic \ous{2}{allows the
programmable core to offload common memory controller operations}. 
\atb{6}{For example,} receive and respond to main memory requests~\circled{6} and
\ous{2}{transfer commands and data between} DRAM Bender \ous{2}{and hardware
buffers}.

\head{Command Buffer~\circled{7}} The command buffer stores and accumulates
multiple DRAM commands before they are issued to the DRAM chip in a
\emph{timing-preserving} batch. The delay between each DRAM command in a batch
is executed exactly as intended by the \X{} user.

\head{Readback Buffer~\circled{8}} Data read from DRAM and returned to \X{} by
DRAM Bender is placed into this buffer. The programmable core receives the
necessary data from the readback buffer for the responses of main memory
requests.


\head{Request and Response Buffers~\circled{9}} EasyTile implements two buffers
\om{6}{(i.e., Incoming Req FIFO and Outgoing Req FIFO)}
that store main memory requests and their responses until they are processed.
The Tile Control Logic automatically inserts incoming main memory
requests and sends responses as they become valid.
\subsection{EasyAPI}
\label{sec:easyapi}

\X{} contains various hardware and software components. \ous{2}{A programmer
orchestrates these components} to evaluate DRAM techniques. We introduce the
user-friendly EasyAPI to \ous{2}{simplify development with} \X{}.
Table~\ref{tbl:apigeneral} shows \ous{2}{some of the most} commonly used
\ous{2}{library} functions of \X{}. We categorize EasyAPI's features into two
categories: hardware (top) and software (bottom).

\head{Hardware Abstraction Library}
The software memory controller interfaces with various hardware modules to
update the system state. For example, a conventional memory controller (e.g.,
FR-FCFS~\cite{zuravleff1997frfcfs, rixner2000memory}) implementation in \X{}
transfers requests between the hardware buffers and the software memory or
updates the time scaling status to emulate the \atb{6}{clock} frequency
constraints requested by the user. \ous{2}{Hardware \atb{6}{abstraction} library
allows} easy interaction with the underlying hardware modules.

\head{Software Library}
The software memory controller implements i) DRAM techniques and ii) \ous{2}{a
memory controller to serve} requests by operating the DRAM module. The
\ous{2}{programmer} may only want to evaluate \ous{2}{a} DRAM technique or
require minimal changes to \ous{2}{an existing} memory controller
\ous{2}{design}. \om{6}{The} \ous{2}{software library streamlines} the design of
these two components \ous{2}{by providing} several DRAM techniques (e.g.,
RowClone~\cite{seshadri2013rowclone}, $t_{RCD}$
reduction~\cite{kim2018solar,wang2018reducing,yaglikci2022hira,
luo2020clrdram,das2018vrldram,hassan2016chargecache,lee2013tiered,lee2015adaptive})
and memory controller (e.g., FCFS, FR-FCFS) implementations.

\section{Time Scaling Validation}
\label{sec:studyconfig}

\head{Simulation}
We validate \X{}'s time scaling against an RTL reference system that is
simulated at a 1 GHz system clock frequency (i.e., same clock for processor,
busses, and caches). The reference system implements \X{}'s memory controller in
hardware, i.e., the reference system's memory controller takes the same
scheduling decisions \om{6}{as} \X{}'s memory controller but does \emph{not}
require time scaling to satisfy target system configuration memory scheduling
latency requirements. We configure \X{} to time scale a processor simulated at
100 MHz clock frequency to 1 GHz. \mirev{We evaluate \param{28} benchmarks from
PolyBench~\cite{pouchet2012polybench} and the memory read latency benchmark from
lmbench suite~\cite{mcvoy1996lmbench}}.\footnote{\mirev{We open-source our
validation reference system and microbenchmarks to enable reproducibility at
\url{https://github.com/CMU-SAFARI/EasyDRAM}}.} Our validation against a 1 GHz
RTL reference system shows that \X{}’s time scaling has low average ($<$0.1\%)
and maximum ($<$1\%) execution time and memory latency inaccuracy
\atb{6}{(measured as the difference in execution time and memory latency between
the RTL reference system and \X{})} across \param{29} evaluated microbenchmarks.
\atb{6}{We attribute the small difference in execution time and memory latency
between the evaluated systems to time scaling.}

\begin{figure}[!h]
\vspace{2mm}
\centering
\includegraphics[width=\columnwidth]{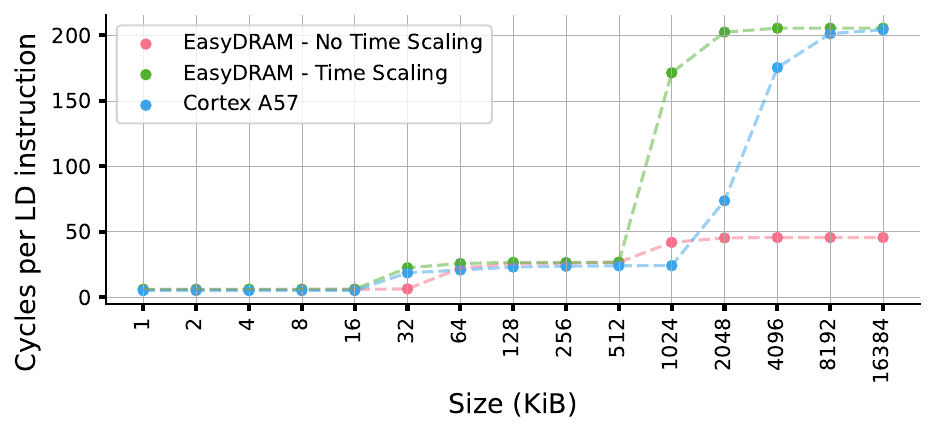}
\caption{Average cycles per load \om{6}{instruction measured} for increasing
\atb{6}{lmbench input sizes} on a real ARM Cortex A57, EasyDRAM - \atb{6}{No
Time Scaling,} and EasyDRAM - Time Scaling}
\label{fig:booma57}
\end{figure}

\head{Real System}
\ous{1}{We extend Chipyard v1.8.1 with our implementation of \X{}. We use a
Xilinx VCU108~\cite{vcu108} board with a DDR4 main memory.\footnote{\X{} can
support additional boards with new top-level FPGA wrappers (referred to as an
fpga-shell or a harness~\cite{amid2020chipyard}) for Chipyard. The development
effort needed to port EasyDRAM to a different FPGA board with similar I/O
capability is very small as all EasyDRAM components are reusable across such
FPGA boards. A user interested in porting EasyDRAM to a different FPGA board
\emph{only} needs to make the correct top-level design I/O to FPGA pin
assignments for their board.} For our experiments, we configure \X{} to target
an NVIDIA Jetson Nano SoC~\cite{jetsonnano}}. We adjust both \X{} and BOOM core
parameters to mirror the corresponding ARM Cortex A57 CPU and memory hierarchy.
Figure~\ref{fig:booma57} shows the latency profile in terms of processor cycles
of the real board \ous{0}{(Cortex A57)} and two versions of \X{}: before
(EasyDRAM - \om{6}{No Time Scaling}) and after (EasyDRAM - Time Scaling)
calibration. We obtain the latency profiles after executing a microbenchmark
similar to the lmbench suite's memory read latency
benchmark~\cite{mcvoy1996lmbench}. The figure indicates the access latency for
reading data from the L1D cache, L2 cache, and main memory. We note that
\atb{6}{the \X{} system has a 512 KiB L2 cache whereas Jetson Nano has a 2 MiB
L2 cache.}

\atbcrcomment{6}{distance between blue and green is due to l2d cache size
discrepancy is this now clear?} 
From \figref{fig:booma57} we observe that,
EasyDRAM - \atb{6}{No Time Scaling} (i.e., a system emulation that does
\emph{not} faithfully model a modern processor) has significantly lower main
memory access latency compared to a real system (i.e., a system with a Cortex
A57 processor). This is because the less-capable processor has slower clock
speed and, thus, fewer \atb{6}{processor clock} cycles \atb{6}{pass} while the
request is served from the main memory. In contrast, EasyDRAM - Time Scaling
(i.e., a system emulation that mimics a board with Cortex A57) has a similar
memory latency \atb{6}{profile} compared to a real system. We conclude that
\atb{6}{\X{} more accurately models a modern system by faithfully emulating main
memory accesses and their impact on execution time than existing FPGA-based
platforms that do \emph{not} faithfully emulate main memory accesses.}

\section{Case Study: Realistic Performance Evaluation of RowClone}

RowClone~\cite{seshadri2013rowclone} is a DRAM technique that copies a row's
content to another row. A state-of-the-art end-to-end processing-in-memory
system evaluation framework (PiDRAM~\cite{olgun2022pidram}) \atb{6}{implements
and evaluates RowClone end-to-end, demonstrating significant system performance
benefits (e.g., 14.6$\times{}$) for RowClone over traditional CPU-based copy operations}. PiDRAM
evaluates RowClone \atb{6}{in a system with a simple in-order processor clocked
at \SI{50}{\mega\hertz} and a 800 MT/s real DDR3 chip. We hypothesize that
PiDRAM's significant system performance benefits for RowClone are attributed to
the disparity between processor and DRAM clock frequency. \atb{6}{To assess
RowClone's performance benefits while faithfully capturing the timing behavior
of a modeled real Cortex A57-based system (see~\secref{sec:studyconfig}), we}
revisit PiDRAM's RowClone evaluation and evaluate RowClone end-to-end using
\X{}.}


\subsection{\atb{6}{RowClone Memory Allocation Constraints}}

\begin{figure}[ht]
\centering
\includegraphics[width=\columnwidth]{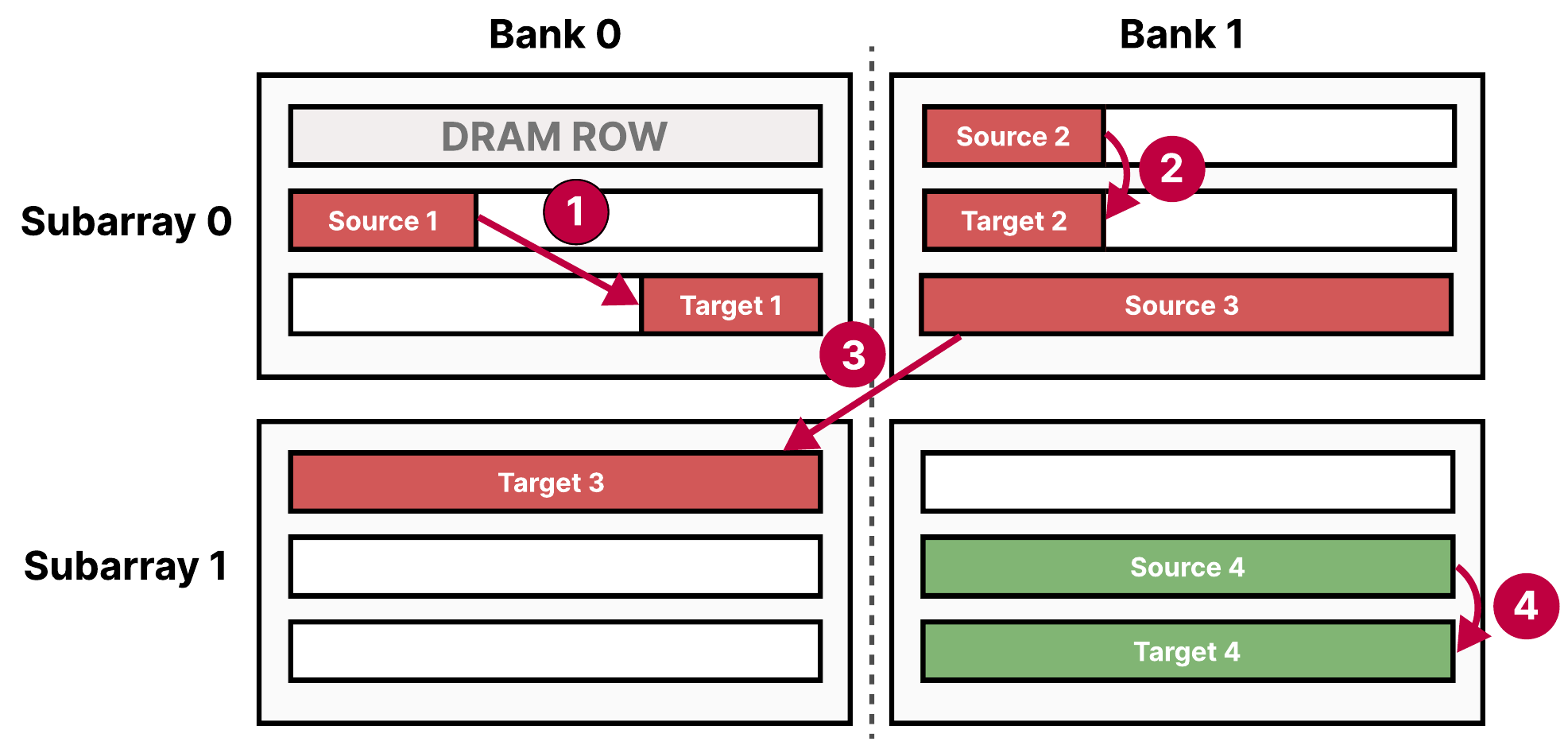}
\caption{\atb{6}{Fast Parallel Mode (FPM) RowClone data allocation constraints}
in a \atb{6}{real COTS} DRAM chip with 2 banks and 2 subarrays. Only operation
\capcircledres{4} can succeed as its operands satisfy all constraints.
Reproduced from~\cite{olgun2022pidram}.}
\label{fig:rcexplain}
\end{figure}

\atb{6}{Fast Parallel Mode (FPM) RowClone~\cite{seshadri2013rowclone} can be
performed in real COTS DRAM chips to move data at row granularity inside a DRAM
subarray~\cite{olgun2022pidram,gao2019computedram,yuksel2024simultaneous}.}
Given that \atb{6}{FPM} RowClone is an intra-subarray operation \atb{6}{and
therefore} the source and target \atb{6}{operands of a RowClone operation} must
satisfy \param{four} constraints\atb{6}{, as depicted
in~\figref{fig:rcexplain},} 1) data must be aligned to row boundaries (alignment
problem \circledres{1}), 2) data size must be a multiple of the DRAM row size
(granularity problem \circledres{2}), 3) both the source and target DRAM rows
must share the same DRAM subarray (mapping problem \circledres{3}), and 4) data
must be resident in memory (\om{6}{coherence} problem).

These restrictions require careful data \atb{6}{allocation} to utilize the full
potential of \atb{6}{our implementation of} RowClone. We demonstrate how
\ous{2}{programmers} can \atb{6}{relatively} easily satisfy these constraints
using \X{}.\atbcrcomment{6}{should be as easy as PiDRAM (no easier than or no
harder than). What do we say here?}

\head{Alignment Problem}
\ous{1}{\atb{6}{The data allocation algorithm (that runs on} the processor) reserves
whole DRAM rows for RowClone operations.} EasyAPI includes a set of
physical-to-DRAM address mappers that allow both the \ous{1}{processor
and \smc{}} to convert an address into a \om{6}{<bank, row, column>} triplet,
and vice versa.

\head{Granularity Problem}
\atb{6}{The data allocation algorithm} selects bank and row pairs that
accommodate the \ous{2}{required} \om{6}{copy} size. The algorithm could choose
to interleave its rows across multiple banks to \mirev{improve}
throughput\om{6}{;} we leave the exploration and optimization \atb{6}{such as
bank interleaving} for \ous{2}{future work}.

\head{Mapping Problem}
\atb{6}{We determine a source and target row address pair to be clonable by
testing using 1000 RowClone copy operations from the source to the target row
(as described in PiDRAM~\cite{olgun2022pidram}). A RowClone copy operation
succeeds if the source row's data equals the target row's data after the
RowClone copy operation. Clonable address pairs are those that never fail across
1000 RowClone copy operations.}

\head{Coherence Problem}
Our processors do \emph{not} implement an instruction to \ous{2}{evict or flush}
cache lines. We provide the \ous{2}{programmer with a memory-mapped} register
that \ous{2}{flushes a} target cache line \ous{2}{to main memory}. This feature
is used to flush the content to DRAM before initiating a RowClone operation and
ensuring data \om{6}{coherence} across the system.

\head{\mirev{Source and Target Row Allocation}} \mirev{A workload can perform a
bulk copy or initialization operation that \emph{cannot} be served by a single
source and destination row address pair \atb{6}{(e.g., if the size of the
operation exceeds the DRAM row size)}. Thus, we distribute a bulk copy
operation's source and destination rows across multiple subarrays. To
accommodate bulk data initialization operations that span multiple subarrays, we
allocate one source row in each subarray. If a RowClone operation between a
source row and a destination row \emph{cannot} be done successfully \atb{6}{(see
Mapping Problem)}, \X{} falls back to regular CPU-based read and write
operations (i.e., uses load and store instructions).}

\subsection{Evaluation and Results}

\mirev{We implement and evaluate RowClone using \X{} and a state-of-the-art
cycle-level software memory simulator, Ramulator 2.0~\cite{luo2023ramulator2,
ramulator2github}. We evaluate two \X{} configurations}. 1)
\ous{0}{\emph{EasyDRAM - \atb{6}{No Time Scaling}}, where we disable time
scaling and 2) \ous{0}{\emph{EasyDRAM - Time Scaling}}, where processors are
emulated as Cortex A57 cores}. \atb{6}{\emph{EasyDRAM - No Time Scaling}}
configuration results allow us to compare the accuracy of time scaling to prior
evaluation methodologies (e.g.,~\cite{olgun2022pidram}). \atb{6}{We configure
\emph{\X{} - No Time Scaling} to resemble the computing system modeled in
PiDRAM~\cite{olgun2022pidram}. \emph{\X{} - No Time Scaling} and PiDRAM model
the same system except that PiDRAM does \emph{not} implement an L2 cache and
\X{} has a 512 KiB L2 cache.} \mirev{We configure Ramulator 2.0 to faithfully
model \X{}'s memory system.}\mirev{\footnote{\mirev{\atb{6}{\X{}'s memory system
consists of 1)~a 512 KiB 8-way last-level cache and 2)~a single-channel and
single-rank of DDR4 chips at 1333 MT/s speed rate with 4 bank groups, 4 banks,
and 32K rows.} The processor system model \atb{6}{that we use in Ramulator 2.0}
(a simple out-of-order core and a last-level cache) significantly differs from
that of \X{}'s real processor system implementation.}}}

\mirev{We execute two workloads}: \textit{Copy} and \textit{Init}. Both programs
take an argument \textit{N}, where \textit{Copy} replicates an \textit{N}-byte
\atb{0}{source} array into another \textit{N}-byte \atb{0}{destination} array,
and \textit{Init} initializes an \textit{N}-byte array with \atb{0}{a}
predetermined \om{6}{data} pattern. Each program comes \atb{6}{with} two
variants: 1) \textit{CPU-copy} duplicates or initializes data by using load and
store \atb{0}{instructions}, and 2) \textit{RowClone} uses \atb{0}{in-DRAM copy}
operations to perform copy and initialization \atb{6}{(and falls back to
CPU-based ld/st instructions, i.e., to \textit{CPU-copy}, if a RowClone copy
operation from the source row to the target row \emph{cannot} be reliably
performed)}.

\mirev{We evaluate these two workloads in \param{two} settings.} In the first
setting, \emph{RowClone \om{6}{- No Flush}}, the source array's data is already
present in DRAM. In the second setting, \emph{RowClone \om{6}{- CLFLUSH}}, the
source array's data \mirev{has cached copies in the processor that must be
written back to DRAM}. The first \ous{0}{setting} demonstrates RowClone's
\atb{6}{best} performance by eliminating any setup overhead associated with the
operation (e.g., flushing caches), and the second \ous{0}{setting} provides a
\atb{6}{worst-case} performance study.\omcomment{6}{Common case is when all cache lines
are in cache. write better}

\begin{figure}[h]
\centering
\includegraphics[width=\columnwidth]{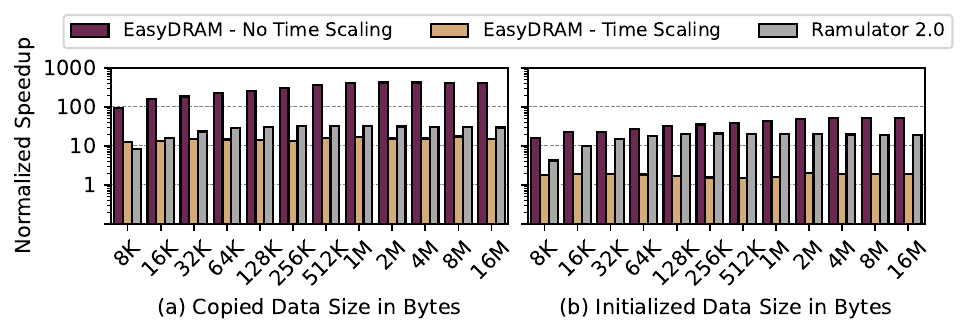}
\caption{\atb{0}{RowClone \atb{6}{- No Flush} \ous{0}{Copy (a) and Init (b)}
execution time speedup normalized to a baseline system that \ous{0}{copies or
initializes} the data with \atb{6}{CPU-based} load and store instructions}}
\label{fig:rowclonenoflush}
\end{figure}

\head{RowClone \om{6}{- No Flush}} \figref{fig:rowclonenoflush} presents the
execution time of RowClone - No Flush for Copy (a) and Init (b) workloads on
increasing array sizes, normalized to each configuration's CPU baseline that
copies or initializes data with load and store instructions. \ous{0}{The axes
respectively show} the \om{6}{input data size} in \ous{0}{bytes (x axis) and the
execution time speedup \om{6}{in} normalized execution time (y axis)}.

\omcomment{6}{Why init so low???}
We make \param{three} observations. First, without time scaling, RowClone Copy
and Init improve the average (maximum) system throughput across different copied
data sizes by 306.7x (423.1x) and 36.7x (51.3x), respectively. Second, with time
scaling average (maximum) \om{6}{RowClone} improvements for Copy and Init
\om{6}{are} 15.0x (17.4x) and 1.8x (2.0x),\atbcrcomment{6}{Say you want to
initialize 4 subarrays worth of DRAM rows. EasyDRAM allocates one source row in
4 subarrays for the init data. Then we copy from this source row to all target
rows that we know that can successfully perform the rowclone operation. For the
remaining rows, we use cpu ld/st. This is not what happens in Ramulator. In
Ramulator we just assume every copy operation is successful. I expanded on this
in the next footnote} respectively. \ous{0}{We attribute the significant
decrease in RowClone's benefits \om{6}{with} time scaling \atb{6}{to the higher
performance of the CPU baseline enabled by \X{}'s ability to more faithfully
model a real system.}} \mirev{Third, with Ramulator 2.0, RowClone Copy and Init
improve the average (maximum) system throughput across all tested data sizes by
27.2x (33.0x) and 17.3x (21.0x), respectively. \mirev{All source and destination
row pairs can successfully perform RowClone operations in Ramulator 2.0
simulations.} We attribute Ramulator 2.0's higher speedup than \emph{\X{} - Time
Scaling} for Init to the lack of real DRAM characterization in software
simulation. Software simulation fails to capture the overhead of falling back to
CPU initialization for rows that \emph{cannot} be initialized with
RowClone}.\footnote{\atb{6}{If the target row of an init operation \emph{cannot}
be successfully initialized using RowClone operations from the source row in a
subarray, \X{} falls back to using CPU-based ld/st operations. We do not model
this behavior in Ramulator 2.0. Our Ramulator 2.0 simulations assume that all
target rows can be successfully initialized using RowClone operations.}}

\begin{figure}[h]
\centering
\includegraphics[width=\columnwidth]{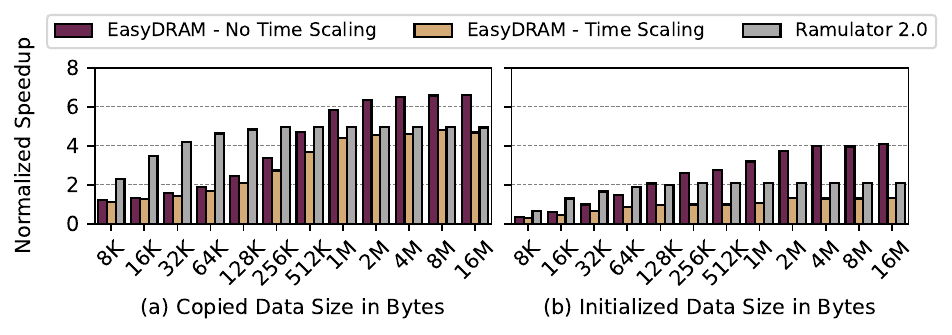}
\caption{\atb{0}{RowClone \atb{6}{- CLFLUSH} \ous{0}{Copy (a) and Init (b)}
speedup normalized to a baseline system that \ous{0}{copies or initializes} the
data with \atb{6}{CPU-based} load and store instructions}}
\label{fig:rcscalecopy}
\end{figure}

\head{RowClone \om{6}{- CLFLUSH}} \figref{fig:rcscalecopy} \ous{0}{presents} the
normalized execution time speedup \ous{0}{of} RowClone \om{6}{- CLFLUSH} for
\om{6}{C}opy \ous{0}{(a)} and \om{6}{I}nit \ous{0}{(b)} operations on increasing array
sizes. \ous{0}{The axes respectively show} the \om{6}{input data size} in
\ous{0}{bytes (x axis) and the execution time speedup (y axis) normalized to
each configuration's CPU baseline without using RowClone}. 

We make \mirev{five} observations. First, with and without time scaling,
RowClone Copy improves the average (maximum) system throughput across all data
sizes by 4.04x (6.62x) and 3.1x (4.83x), respectively. Second, \mirev{with} and
without time scaling, RowClone Init improves the average (maximum) system
throughput for data sizes above 256KB by 1.19x (1.34x) and above 32KB by 3.09x
(4.08x), respectively. \mirev{RowClone Init with and without time scaling
degrade system performance for array sizes that are $\leq{}\!$ 256 KB and 32 KB,
respectively.}
Third, when compared to RowClone \atb{6}{- No Flush}
\mirev{(\figref{fig:rowclonenoflush})}, the copy and initialize operations on
small source data array sizes suffer significantly as \atb{6}{our implementation
of RowClone flushes dirty cache blocks of source DRAM rows (and invalidates
clean cache blocks of the target DRAM rows) for coherence}. Coherence overheads
are less evident in large data sizes because dirty block flush operations
overlap with more data accesses. Fourth, as the source data array size
increases, RowClone \atb{6}{- CLFLUSH} system performance benefits similarly
increase. \mirev{Five, Ramulator 2.0 yields a different speedup profile than
\X{}. We attribute the difference to the difference in processor systems modeled
by Ramulator 2.0 and \X{}.}

Based on our experiments, we conclude that evaluation frameworks that do
\emph{not} faithfully model a \ous{2}{modern} \mirev{processor} (e.g., $>$1 GHz
clock frequency) report relatively high benefits in favor of DRAM techniques
with large error margins (e.g., $>$20x). We note that RowClone still
\atb{6}{significantly} improves system performance when compared to a
\ous{2}{modern} CPU baseline, especially at large source data array sizes (e.g.,
$>$\atb{6}{64 KiB}).\atbcrcomment{6}{not sure what to fix,
made array size 64 KiB where RC-copy exceeds 2x normalized speedup}
\section{Case Study: \om{6}{DRAM} Access Latency Reduction}

Several techniques have been proposed in prior
work~\cite{kim2018solar,wang2018reducing,yaglikci2022hira,luo2020clrdram,
das2018vrldram,hassan2016chargecache,lee2013tiered,lee2015adaptive,chandrasekar2014exploiting,
chang2016understanding,lee2017design} to improve the main memory latency. To
understand the system-wide benefits of these approaches we implement and
evaluate row-to-column command delay ($t_{RCD}$) timing parameter reduction with
\X{} \atb{6}{(based on especially~\cite{kim2018solar})}.

\subsection{$t_{RCD}$ Reduction Implementation}
\label{sec:solarcharacterization}

We implement $t_{RCD}$ reduction in two stages. First, we profile DRAM to
determine the rows that serve correct data under reduced $t_{RCD}$ values
\atb{0}{(strong DRAM rows)}. Second, we extend our memory controller to
intelligently access strong DRAM rows with the reduced \om{6}{$t_{RCD}$} timing
parameter to reduce DRAM access latency.

\head{DRAM Characterization}
\ous{1}{We perform our DRAM characterization using both the processor
and the \smc{}}. \ous{1}{We extend the \smc{} to support a new type of request
called \emph{profiling request} that tests a cache line for a given $t_{RCD}$
value}. The \smc{} \ous{1}{serves a profiling request} in \param{three} steps:
1) initialize the target cache line with a known data pattern, 2) access the
target cache line using the requested $t_{RCD}$ value, and 3) report to the
processor whether the reduced value resulted in correct access.
\ous{1}{During execution, the \smc{} serves both main memory and profiling
requests and the} processor iterates over banks, cache lines, and rows
\ous{1}{to test} while keeping track of the locations that can \ous{1}{be
accessed reliably} under reduced $t_{RCD}$ values.

\begin{figure}[h]
\centering
\includegraphics[width=\columnwidth]{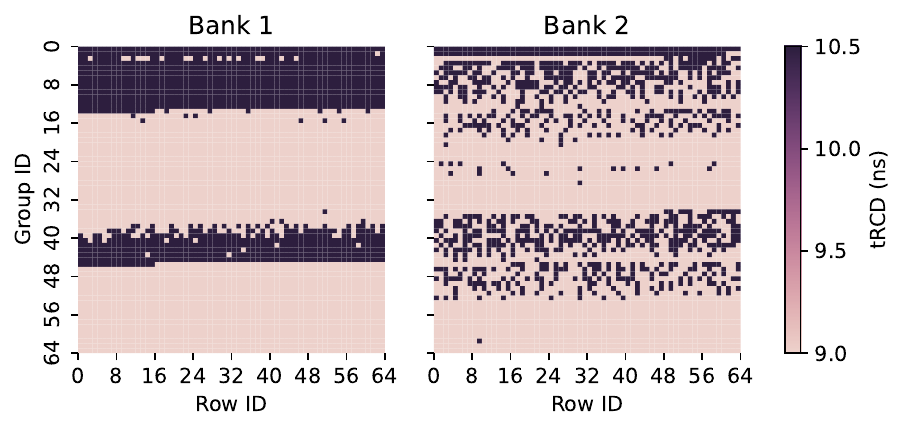}
\caption{Minimum \ous{2}{reliable} $t_{RCD}$ \ous{2}{of rows across two banks}}
\label{fig:solarchar}
\end{figure}

Figure~\ref{fig:solarchar} \atb{6}{shows} a \ous{1}{profiling heatmap for the} first
two banks across 4K rows per bank. \atb{6}{Nominal $t_{RCD}$ for the DRAM module we test
is \SI{13.5}{\nano\second}~\cite{ddr4chipundertest}.}

\atb{6}{We make three major observations from this figure. First, all cache
lines can reliably operate at $t_{RCD}$ values lower than the nominal $t_{RCD}$
value (i.e., all cells are stronger than what the manufacturer recommends but
some rows are even stronger). This finding is consistent with prior
studies~\cite{kim2018solar,wang2018reducing}.}\atbcrcomment{6}{better now?}
\atb{6}{We consider a cache line \emph{strong} or \emph{weak},
respectively, if the cache line can be reliably accessed using
$\leq{}\!$\SI{9.0}{\nano\second} or $>\!$ \SI{9.0}{\nano\second} $t_{RCD}$.}
\atb{6}{Second}, the percentage of strong DRAM \atb{6}{cache lines} (84.5\%)
significantly exceeds the weak \atb{6}{cache lines} (15.5\%). \atb{6}{Third},
weak \atb{6}{cache lines} are clustered within specific banks and areas. We
conclude that \om{6}{all rows can operate faster than nominal $t_{RCD}$ but a
vast majority of rows can operate even faster} to reduce memory \atb{6}{access}
latency.

\subsection{Scheduler Implementation}
\label{sec:bloomfilter}

Storing the minimum $t_{RCD}$ value of all cache lines is \emph{not} scalable as
DRAM capacity increases \ous{0}{(e.g., $>$64 GiB)}. We introduce a two-part
strategy to address this challenge. First, we identify the weakest cache line in
each row and use its $t_{RCD}$ value \atb{6}{as the $t_{RCD}$ value of every
cache line in that row}. Second, we \ous{1}{implement a} \om{6}{B}loom
filter~\cite{bloom70spacetime}\om{6}{, similarly to RAIDR~\cite{raidr},}
\ous{1}{in the \smc{} that tracks} weak DRAM rows. We use weak rows as keys for
our Bloom filter \atb{6}{such that a false positive does \emph{not} cause a
reduced $t_{RCD}$ access to a weak row}. We generate the Bloom filter and the
hash functions on the host machine and load them \ous{0}{to} \smc{} before
emulation begins. \ous{0}{Each time a DRAM row is opened}, \smc{} checks
\atb{6}{the Bloom} filter to determine the appropriate $t_{RCD}$ value
\ous{0}{to use}.

\subsection{Evaluation and Results}

\mirev{We implement and evaluate $t_{RCD}$ reduction using \emph{\X{} \atb{6}{-
Time Scaling}} and Ramulator 2.0~\cite{luo2023ramulator2, ramulator2github}. For
\X{}, we leverage the fast emulation capabilities of \X{} and run workloads
\atb{6}{to completion} to study system behavior. For Ramulator 2.0, we generate
traces of workloads and simulate each workload for 500M instructions. We
evaluate workloads from the large dataset configuration of
PolyBench~\cite{pouchet2012polybench}, a suite of benchmark kernels widely used
in high-performance computing.} \atb{6}{We use the $t_{RCD}$ value identified by
the two-step process 1)~real DRAM chip
characterization~(\secref{sec:solarcharacterization}) and 2)~Bloom filter
initialization~(\secref{sec:bloomfilter}) for each DRAM row.}

\begin{figure}[h]
\centering
\includegraphics[width=\columnwidth]{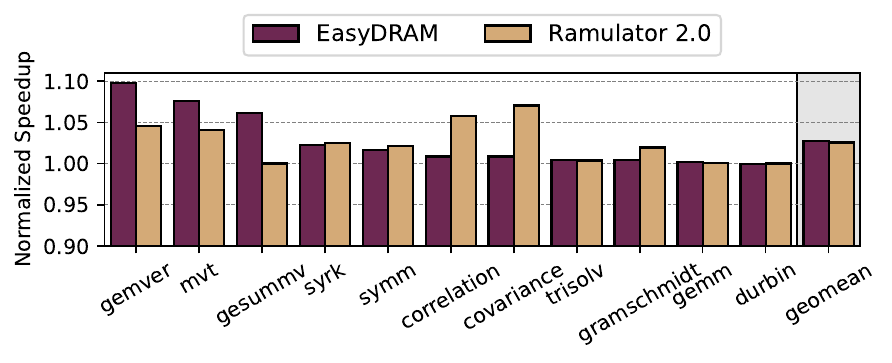}
\caption{Execution time speedup \om{6}{with} $t_{RCD}$ reduction}
\label{fig:polytrcd}
\end{figure}

\figref{fig:polytrcd} presents the execution time normalized to the baseline
system \emph{without} $t_{RCD}$ reduction. \atb{6}{The y-axis shows execution
time speedup normalized to a system that uses nominal $t_{RCD}$
(\SI{13.5}{\nano\second}).}

\omcomment{6}{Unclear what trace reduction is used for what rows. Unclear what
data is allocated where. Many important descriptions in evaluation are missing.
Unclear what the baseline trace value is. Need to clarify all earluier before
presenting results. } \mirev{We make two observations from
\figref{fig:polytrcd}}. \mirev{First, with \X{}}, $t_{RCD}$ reduction improves
performance on average (maximum) by 2.75\% (9.76\%).\footnote{\atb{6}{The
evaluated workloads have last-level cache misses per kilo processor cycles of
\emph{only} 2.2 on average (i.e., the workloads are \emph{not} memory
intensive), thus, the performance improvement provided by \X{} with $t_{RCD}$
reduction is reasonable and we expect it to increase with the memory intensity
of workloads.}} \mirev{Second, with Ramulator 2.0, $t_{RCD}$ reduction improves
performance on average (maximum) by 2.58\% (7.04\%).} \mirev{\X{} and Ramulator
2.0 can yield different results for individual workloads (e.g., correlation)
because Ramulator 2.0 1)~simulates only a part of the workload and 2)~\atb{6}{is
configured to simulate} a simple out-of-order processor core model whereas \X{}
executes the \atb{6}{workload to completion} and implements a real out-of-order
processor core (BOOM~\cite{celio2018thesis}).}

\begin{figure}[h]
\centering
\includegraphics[width=\columnwidth]{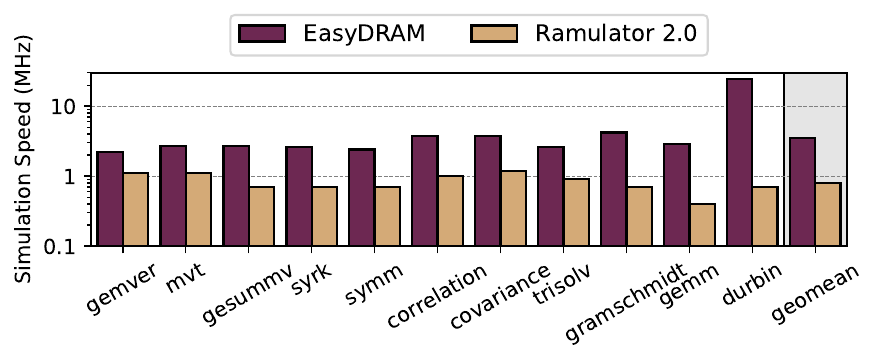}
\caption{\X{}'s \atb{6}{simulation} speed across benchmarks}
\label{fig:polysimspeed}
\end{figure}

\mirev{We study the evaluation speed of \X{} and Ramulator 2.0.}
\figref{fig:polysimspeed} presents the evaluation speed of \X{} \mirev{and
Ramulator 2.0} across benchmarks. The x-axis shows the evaluated workload, and
the y-axis shows the \atb{6}{simulation} speed \atb{6}{(computed as simulated
processor cycles/time)} in MHz.\atbcrcomment{6}{durbin is not memory intensive
so EasyDRAM makes very fast progress.}

\mirev{From \figref{fig:polysimspeed} we observe that \X{}'s average (maximum)
\atb{6}{simulation} speed is \atb{6}{5.9$\times{}$ (20.3$\times{}$)} faster than
Ramulator~2.0}. \atb{6}{\X{}'s simulation speed improvement over Ramulator 2.0
increases with reducing memory intensity of the evaluated workload (e.g., durbin
has a last-level cache misses per kilo processor cycles of \emph{only}
0.01).}\omcomment{6}{is this normaliz per instruction?
unclear.} We conclude that
\X{} allows fast full-system evaluation of DRAM techniques using \om{6}{real}
DRAM chips.

\section{Related Work}

\om{6}{To our knowledge, \X{} is the first work to develop an FPGA-based
end-to-end framework for accurate and rapid evaluation of DRAM techniques using
real DRAM chips.} We have already provided qualitative comparisons to the most
closely related work that could be used to evaluate DRAM
techniques~\cite{binkert2011gem5,kim2016ramulator,luo2023ramulator2,forlin2022sim2pim,power2014gem5,biancolin2019fased,
karandikar2018firesim,hassan2017softmc,olgun2022pidram,olgun2023drambender,mosanu2022pimulator,mosanu2023freezetime}. In
this section, we present other related work.

\head{Programmable memory controllers (e.g.,~\cite{goossens2013reconfigurable,
bojnordi2012pardis})} Programmable memory controllers allow dynamic
specialization of the memory controller for any running application. \om{6}{Such
programmable controllers as in}~\cite{goossens2013reconfigurable,
bojnordi2012pardis} \atb{6}{have different goals than \X{} and} are \emph{not}
\om{6}{suitable for} evaluating DRAM techniques for three reasons. First, these
works have limited programmability because they aim to achieve hardware design
timing closure at very high clock speeds. Second, similar to commercial systems,
these works do \emph{not} accommodate easy modifications to evaluate varying
system configurations. Third, these works require users to write programs using
their \om{6}{own} specialized instruction set architecture.\footnote{ A
compiler could ease the programmability of existing programmable memory
controllers. However, \X{} is superior to a compiler for programmable memory
controllers (e.g.,~\cite{goossens2013reconfigurable, bojnordi2012pardis}) in the
evaluation of DRAM techniques due to two reasons. First, \X{} provides a
flexible and expressive software-defined memory controller that can be used to
rapidly evaluate varying DRAM techniques. In contrast, a compiler would
\emph{not} improve the expressiveness (e.g., adding new operations such as
RowClone) of existing programmable memory controllers to a point where they
could match \X{}’s. Second, without time scaling, a programmable memory
controller is subject to the problems we describe in~\secref{sec:motivation},
which prevents accurate evaluation of new and existing DRAM techniques.} In
contrast, \X{} 1) allows every aspect of the memory controller to be customized,
2) accommodates easy configuration of multiple components of the system (e.g.,
cache sizes and processor capabilities), and 3) can be programmed with a common
high-level language (i.e., C++).

\head{MEG~\cite{zhang2020MEG}} MEG is a system emulation infrastructure for
near-data processing that uses High-Bandwidth Memory (HBM). MEG implements a
soft core as a processor in the FPGA and is designed to efficiently retrieve
data from main memory to perform computation in the processor. MEG's goal is
\emph{not} to allow for rapid and accurate evaluation of DRAM techniques.
However, parts of MEG's system design can be integrated into \X{} to allow for
rapid and accurate evaluation of DRAM techniques using real HBM chips.

\head{ComputeDRAM~\cite{gao2019computedram}} ComputeDRAM demonstrates
\om{6}{RowClone~\cite{seshadri2013rowclone} and bulk bitwise AND, OR,
MAJ~\cite{seshadri2015fast,seshadri2017ambit} operations} in real off-the-shelf
DDR3 DRAM chips. ComputeDRAM describes a software framework built around
SoftMC~\cite{hassan2017softmc,softmcgithub} to estimate in-memory computation
throughput. This framework is subject to the limitations of
SoftMC~\cite{hassan2017softmc} that are described in
Section~\ref{sec:introduction}.

\head{Clock \om{6}{Frequency Emulation} on FPGAs} \atbcrcomment{6}{Pretty sure
after skimming through FAST et al. that FAST is not designed for this
problem}Previous FPGA-based
simulators~\cite{karandikar2018firesim,biancolin2019fased} emulate different
clock frequencies by throttling parts of the system depending on their
\om{6}{targeted} simulation speed. They achieve this by treating all input
and output signals as tokens. A module (e.g., the memory controller) only
advances a time step (i.e., produces an output token) if it can consume one
token from all of its inputs. The token-based method is mainly suited for
hardware-based memory controllers and incurs impractical performance overheads
for a software memory controller because the programmable core inefficiently
\ous{2}{processes} input and output tokens. 
\atb{6}{We develop time scaling
building on the basic idea of throttling parts of the system to emulate different
clock frequencies for software memory controller-based FPGA systems.}

\head{Other \om{6}{Related Works}} Many prior \om{6}{works} propose or
demonstrate \atb{6}{various} DRAM techniques (e.g., processing in
memory~\cite{kim.hpca18,talukder2019exploiting,orosa2021codic,
olgun2021quactrng,kim2019drange,seshadri2013rowclone,gao2019computedram,seshadri2017ambit,
yuksel2024simultaneous,yuksel2024functionally,gao2022frac}, access latency
reduction~\cite{lee2015adaptive, chang2016understanding, chang2017thesis,
lee2017design, kim2018solar}, \atb{6}{retention-aware intelligent
refresh}~\cite{qureshi2015avatar,liu2012raidr,patel2017reaper,rapid}).
None of these works provide a framework that can rapidly and
accurately evaluate DRAM techniques end-to-end. The techniques proposed by these
works can be implemented in \X{} and their performance benefits evaluated,
similar\om{6}{ly} to how we do so for RowClone~\cite{seshadri2013rowclone} and
SolarDRAM~\cite{kim2018solar}.
\section{Conclusion}

\atb{6}{We introduced a new, easy-to-use and extensible FPGA-based framework for
accurate and fast evaluation of DRAM techniques using real DRAM chips end-to-end
in a real system. Compared to state-of-the-art FPGA-based evaluation platforms,
\X{} significantly improves evaluation accuracy. \X{} combines two ideas: 1)~it
enables developers to implement DRAM techniques using a high-level language
(C++) to remove the need for hardware design expertise and 2)~it advances the
FPGA system state in a way that faithfully captures the timing behavior of the
modeled system using time scaling. While effective at improving evaluation
accuracy, \X{} is easy to use in modeling DRAM techniques because using \X{}
does \emph{not} require deep hardware description language expertise. We
believe and hope that \X{} will enable innovative ideas in memory system design
to rapidly come to fruition.}

\section*{\mirev{Acknowledgments}} {\mirev{We thank the anonymous reviewers of
ASPLOS 2024 and DSN 2025 for the encouraging feedback. We thank the SAFARI
Research Group members for valuable feedback and the stimulating scientific and
intellectual environment. We acknowledge the generous gift funding provided by
our industrial partners (especially Google, Huawei, Intel, Microsoft, VMware).
This work was also in part supported by the Google Security and Privacy Research
Award, the Microsoft Swiss Joint Research Center, the ETH Future Computing
Laboratory (EFCL), \atb{6}{Semiconductor Research Corporation (SRC), AI Chip
Center for Emerging Smart Systems (ACCESS), sponsored by InnoHK funding, Hong
Kong SAR, European Union's Horizon programme for research and innovation
[101047160 - BioPIM]} and the F3CAS project (ANR-20-CE25-0010) of the French
National Research Agency.}}

\balance
\bibliographystyle{IEEEtran}
\bibliography{refs}

\end{document}